\documentclass[useAMS,usenatbib,usegraphicx]{mn2e}
\usepackage{times}
\usepackage{amsmath}
\usepackage{amssymb}
\usepackage{mathrsfs}

\newcommand{\msun}{\,\mbox{$\rm M_{\odot}$}}
\newcommand{\mearth}{\,\mbox{$\rm M_{\oplus}$}}

\newcommand{\kms}{\hbox{kms$^{-1}$}}
\newcommand{\ms}{\hbox{ms$^{-1}$}}
\newcommand{\cms}{\hbox{cms$^{-1}$}}
\newcommand{\vsini}{\hbox{$v$\,sin\,$i$}}

\newcommand{\degs}{$\degr$}

\newcommand{\ha}{H$\alpha$}

\newcommand{\eev}{{\sc eev}}
\newcommand{\mitll}{{\sc mitll}}
\newcommand{\harps}{{\sc harps}}

\title[M dwarf radial velocities with ROPS]{Precision radial velocities of 15 M5\,-\,M9 dwarfs}

\author[J.R.~Barnes et al.]
{J.R.~Barnes$^{1}$, 
J.S.~Jenkins$^{2}$,  
H.R.A.~Jones$^{1}$,  
S.V. Jeffers$^{3}$,
P. Rojo$^{2}$,  
P. Arriagada$^{4}$, \newauthor 
A. Jord\'an$^{4}$,  
D. Minniti$^{4,5}$,  
M. Tuomi$^{1,6}$, 
D. Pinfield$^{1}$, 
and G. Anglada-Escud\'{e}$^{7}$ \\
$^{1}$ Centre for Astrophysics Research,. University of Hertfordshire,. College Lane, Hatfield. Herts. AL10 9AB. UK. \\
$^{2}$ Departamento de Astronom\'{i}a, Universidad de Chile, Camino del Observatorio 1515, Las Condes, Santiago. Chile. \\
$^{3}$ Institut f\"{u}r Astrophysik, Georg-August-Universit\"{a}t, Friedrich-Hund-Platz 1, Friedrich-Hund-Platz 1, D-37077 Göttingen. Germany \\
$^{4}$ Instituto de Astrof\'isica, Pontificia Universidad Cat\'olica de Chile, Av.\ Vicu\~na Mackenna 4860, 7820436 Macul, Santiago, Chile \\
$^{5}$ Vatican Observatory, V00120 Vatican City State, Italy \\
$^{6}$ University of Turku, Tuorla Observatory, Department of Physics and Astronomy, V\"ais\"al\"antie 20, FI-21500, Piikki\"o, Finland \\
$^{7}$ Astronomy Unit, School of Mathematical Sciences, Queen Mary, University of London. UK.}

\begin{document}

\date{MNRAS, accepted}

\pagerange{\pageref{firstpage}--\pageref{lastpage}} \pubyear{2010}

\maketitle

\protect\label{firstpage}

\begin{abstract}

We present radial velocity measurements of a sample of M5V\,-\,M9V stars from our Red-Optical Planet Survey, ROPS, operating at \hbox{0.652\,-\,1.025 \micron}. Radial velocities for 15 stars, with r.m.s. precision down to \hbox{2.5 \ms}\ over a week long time scale are achieved using Thorium-Argon reference spectra. We are sensitive to planets with \hbox{$m_p$\,sin\,$i$ $\geq$\,1.5\,M$_\oplus$ (3\,M$_\oplus$ at 2-$\sigma$)} in the classical habitable zone and our observations currently rule out planets with \hbox{$m_p$\,sin\,$i$ $>$\,0.5 M$_{\rm J}$} at 0.03 AU for all our targets. A total of 9 of the 15 targets exhibit \hbox{r.m.s. $<$\,$16$ \ms}, which enables us to rule out the presence of planets with \hbox{$m_p$\,sin\,$i$ $>$\,10 \mearth}\ in \hbox{0.03 AU} orbits.

Since the mean rotation velocity is of order \hbox{8 \kms}\ for an M6V star and \hbox{15 \kms}\ by M9V, we avoid observing only slow rotators that would introduce a bias towards low axial inclination \hbox{(i $\ll$ 90\degs)} systems, which are unfavourable for planet detection. Our targets with the highest \vsini\ values exhibit radial velocities significantly above the photon-noise limited precision, even after accounting for \vsini. We have therefore monitored stellar activity via chromospheric emission from the \ha\ and Ca {\sc ii} infrared triplet lines. A clear trend of log$_{10}$(L$_{\rm H_{\alpha}}$/\,L$_{\rm bol}$) with radial velocity r.m.s. is seen, implying that significant starspot activity is responsible for the observed radial velocity precision floor. The implication that most late M dwarfs are significantly spotted, and hence exhibit time varying line distortions, indicates that observations to detect orbiting planets need strategies to reliably mitigate against the effects of activity induced radial velocity variations.
\end{abstract}

\begin{keywords}
(stars:) planetary systems
stars: activity
stars: atmospheres
stars: spots
techniques: radial velocities
\end{keywords}

\section{Introduction}
\protect\label{section:intro}

Although the solar neighbourhood is dominated by low mass stars, the late M dwarf population has remained largely beyond the reach of optical precision radial velocity surveys. In order to address this major parameter space, dedicated instruments have been proposed that would instead operate at longer wavelengths, at the peak of the energy distribution of low-mass stars \citep{jones08prvs}. Upcoming instruments are now being constructed, and include the Habitable Zone Planet Finder \citep{mahadevan12hzp} and {\sc carmenes}, the Calar Alto high-Resolution search for M dwarfs with Exo-earths with Near-infrared and optical Echelle Spectrometers \citep{quirrenbach2012carmenes}. However, while a number of well established instruments with proven stability at earlier spectral types have also reported precision radial velocities (RV) for early M dwarfs, the {\sc crires} survey \citep{bean10b} and the ROPS survey \citep{barnes12rops} (hereafter B12) have reported precision radial velocities at the $\sim$\,$10$ \ms\ level for late M dwarfs (M6V\,-\,M9V) { with existing instrumentation. \cite{reiners09flare} has also reported \hbox{$\sim$\,10 \ms}\ stability on the flaring M6 dwarf CN Leo.} Working in the infrared K band, Bean et al. (2010) reported 11.7 \ms\ for Proxima Cen, and 5.4 \ms\ after observations were binned together. On the other hand, B12, working in the red-optical (0.62\,-\,0.90 \micron) found that while propagated errors were at the $\sim$\,$10$ \ms\ level, the r.m.s.\ scatter was 16\,-\,35 \ms\ in the most stable targets. Until {\sc crires} is upgraded to a cross-dispersed, multi-order instrument, {\sc uves} has substantially more wavelength coverage with reasonable signal-to-noise from which radial velocities may be derived. {\sc uves} has also already demonstrated \hbox{2\,-\,2.5 \ms}\ precision over 7 yrs working with an I$_2$ cell \citep{zechmeister09uves} in the 5000\,-\,6000 \AA\ range. However, by \hbox{$\sim$\,6500 \AA}, I$_2$ lines become weak (\hbox{$<$ 10 per cent} of the normalised continuum) and are barely visible \hbox{beyond 7000 \AA}. Hence I$_2$ gas cells cannot be used in the red part of the optical, beyond these wavelengths.

The first planet orbiting an M dwarf, was reported by \cite{delfosse98gj876} and \cite{marcy98gj876} nearly a decade after the first low mass companion  to the main sequence star HD 114762 \citep{latham89}, which may be either a brown dwarf or massive planet, depending on the unknown orbital inclination. \hbox{GJ 876 b}, orbiting its parent M4V star in a \hbox{61 day} orbit is a giant planet, which is perhaps not surprising given that close-orbiting companions are the easiest to detect with few epochs of observations using radial velocity techniques. However, while close-orbiting planets have been { predicted} to be relatively common { for early M dwarf samples (see \S \ref{section:hz_occurrence} below)}, only $\sim$ 50 per cent of the M dwarf planets with mass estimates (16 from a total of 31)\footnote{http://exoplanets.org} possess masses $\gtrsim$\,$0.3$ M$_{\rm J}$. The remaining 15 planets have minimum masses implying Super-Earth to Neptune-mass companions. \hbox{GJ 876 b} is only one of four planets so far detected { orbiting} \hbox{GJ 876}, { and in fact two of the planets possess} masses of only 5.8 \mearth\ and 12.5 \mearth. { In addition, amongst the Kepler candidates first reported by \cite{borucki11kepler} and confirmed by a number of authors (\citealt{fabrycky12kepler}; \citealt{steffen13kepler}; \citealt{muirhead12kepler}), 14 planets have been identified with radii $\leq 3$ M$_{\oplus}$, whilst no transiting hot Jupiters have been detected.} To date these findings confirm earlier predictions that Neptune mass and Earth-mass planets are expected in greater numbers in orbit around M stars \citep{ida05}.

\subsection{Rocky planet occurrence rates and the M dwarf habitable zone}
\protect\label{section:hz_occurrence}

\cite{bonfils13mdwarfs} have calculated phase-averaged detection limits for individual stars, which enable the survey efficiency of the \harps\ (High Accuracy Radial velocity Planet Searcher) { early} M dwarf sample to be determined. These detection limits enable corrections to be made for incompleteness, allowing occurrence rates to be estimated. The frequency of HZ planets, $\eta_\oplus$, {\hbox{(where 1 \mearth\ $<$ $m$\,sin\,$i$ $<$ 10 \mearth)} orbiting { early} M dwarfs is found to be 0.36$^{+0.25}_{-0.10}$. From the Kepler sample, \cite{dressing13} estimated $\eta_\oplus$ = \hbox{0.90 $^{+0.04}_{-0.03}$} for M dwarf planets up to \hbox{4 R$_\oplus$}. With revised estimates of habitable zones \citep{kopparapu13hz}, \cite{kopparapu13} used the 95 Kepler planet candidates orbiting 64 low-mass host stars to similarly place conservative estimates of $\eta_\oplus$ = 0.51$^{+0.10}_{-0.20}$ for M dwarf planets with radii in the range \hbox{0.5\,-\,2 R$_\oplus$}. \cite{bonfils13mdwarfs} find that the majority of { \em early} M dwarf planets are clustered in few-day to tens of days orbits, continuing the trend with semi-major axis distribution observed by \citep{currie09}. By extrapolation, we might expect { \em late} M dwarf planets in orbits up to a few 10s of days. 

The centre of the continuous habitable zone for a M6V star is estimated to be $\sim$\,0.045 AU \citep{kopparapu13hz}. Hence, a \hbox{7.5 \mearth}\ planet would induce a \hbox{$K_{*}$ = 10 \ms}\ signal with an \hbox{11.0 day} period. Although \citep{kopparapu13hz} do not make habitable zone estimates for low masses, based on a simple flux and mass scaling, we estimate that an M9V star habitable zone would be centred at \hbox{$\sim$0.023 AU}, with a \hbox{7.5 \mearth}\ planet inducing a \hbox{15.8 \ms}\ signal with a \hbox{4.4 day} period. { Observations spanning a six day period (which we present in this paper)}, thus offer the potential to sample 55 per cent of an M6V habitable zone period, and greater than a complete orbit for an M9V star. { By defining a continuous habitable zone, the range of possible orbital periods for habitable planets are extended. For instance, \cite{kopparapu13hz} define inner moist greenhouse and outer greenhouse limits, that extend the range of periods for an M6V habitable planet from $\sim$\,$6$ days to a maximum of $\sim$\,$17$ days.}

\subsection{ROPS Sample}
\protect\label{section:rops_sample}

Our choice of targets was based a number of factors including visibility and brightness. In order to obtain sufficient S/N in the spectra in exposures limited to no more than 1800 secs we limited the selection to M5\,-\,M9 dwarfs with apparent I band magnitudes $\lesssim 14.5$. A number of stars in common with our initial observations made with the {\sc mike} spectrograph at Magellan Clay (B12) have been retained. Additional targets were selected, ensuring that a range of spectral types were included with low-moderate \vsini\ values. Because M stars, and particularly late M stars on the whole are not effectively spun down, those stars later than M6V tend to be moderate rotators on the whole. \cite{jenkins09mdwarfs} found that M6V stars on average possess \hbox{\vsini\ $\sim$\,$8$ \kms}, whereas this rises to \hbox{$\sim$\,$15$ \kms}\ by M9V. This obviously has important consequences for radial velocity precision, especially if magnetic activity phenomena affect the rotation profiles. { Because moderate rotation is found on average, selecting only the slowest rotating stars with \vsini\ $\leq$\,$5$ \kms\ is likely to bias a target sample to { low axial inclination ($i \ll 90$\degs)} systems (i.e. with rotation axis aligned along the line of sight to the observer), for which detection of planets is less favourable}. In order to characterise the effects of activity for this and future surveys, we included moderate rotators in our sample. The objects were selected for which \vsini\ was on the whole well measured \citep{mohanty03activity,reiners10activity}. In addition, following the procedures { detailed} in \cite{jenkins09mdwarfs}, we have also obtained the first \vsini\ measurements for two of the targets in our sample, GJ 3076 and GJ 3146, as indicated in \hbox{Table \ref{tab:targets}}.

{ In this paper we investigate the methods by which precision radial velocities can be achieved with existing instrumentation, extending the search of optical spectrometers into the 0.65\,-\,1.025 \micron\ wavelength region, where no established simultaneous reference fiducial has been tested.} In section \S \ref{section:wavcalib} we outline our master wavelength calibration procedure. The use of tellurics for wavelength calibration is investigated in \S \ref{section:tellurics} using an analysis similar to that carried out by \cite{figueira10stability} for {\sc harps} observations of G type stars. We derive radial velocities for Proxima Centauri using only telluric lines to enable us to determine the simultaneous wavelength solution. In section \S \ref{section:rops_obs} we present the radial velocity measurement procedures for our ROPS sample, discussing our wavelength calibration procedure (\S \ref{section:rops_thar}), applicable particularly to {\sc uves} observations, before presenting radial velocities for our 15 M5V\,-\,M9V targets from 4 epochs of observations spread over a week-long timescale (\S \ref{section:results}). Finally, we discuss our findings (\S \ref{section:discussion}) and prospects for future observations (\S \ref{section:conclusion}).



\section{Observations}
\protect\label{section:observations_intro}

{ 
In this paper, we utilise observations made during our own observing campaign in 2012 July. We also use data taken from the European Southern Observatory ({\sc eso}) archive.
}
\subsection{ROPS observations with UVES}
\protect\label{section:obs_rops}
We observed 15 M dwarf targets with the Ultraviolet and Visual Echelle Spectrograph {\sc uves} at the 8.2m Very Large Telescope ({\sc vlt, ut2}). Observations were made with a 0.8\arcsec\ slit, which give a resolution of R $\sim$ 54,000. We observed on four half nights spread over a period of six nights in total, on 2012 July 23, 24, 26 \& 29 (UTC). Short orbital periods might be expected by extrapolating the tens of days orbits, found amongst {\em early} M dwarfs \citep{bonfils13mdwarfs}, to the late M dwarf population. Additionally the observing strategy enabled the stability of {\sc uves} and our measurement precision to be characterised on week long timescales. Although {\sc uves} offers the ability to simultaneously record observations at shorter and longer wavelengths, we opted to make observations in the red arm only since mid to late-M stars output little flux short of 6000 \AA. In B12, we found the ratio of flux in the 7000\,-\,9000 \AA\ region compared with the 5000\,-\,7000 \AA\ region to be 11.5 and 19 for M5.5V and M9V spectra respectively. This estimate included the throughput of the 6.5m Magellan Clay and {\sc mike} spectrograph. 

Working in the red-optical (i.e. 0.6\,-1.0 \micron) poses a particular challenge in that there no currently operating \'{e}chelle spectrometers coupled with 8m class telescopes that offer simultaneous calibration. Regular wavelength observations for calibration are crucial if precisions of order \ms\ are to be achieved from high resolution radial velocity information. Although {\sc uves} possesses an iodine cell, the absorption lines of I$_{2}$ do not extend far above 6500 \AA, and are already very weak, with line depths of only a few per cent of the normalised continuum. We have therefore opted to utilise near-simultaneous observations of Thorium-Argon (ThAr) arc lamp lines, coupled with the relative stability of {\sc uves} in order to achieve sub-\ms\ precision on our target population of late M stars. Since ThAr lamps exhibit many lines for calibration, and are generally always available by default with \'{e}chelle spectrometers working at optical wavelengths, we made regular observations with the comparison lamp available with {\sc uves}. A calibration was included in the observing block associated with each observed target and was taken immediately after each science frame. Further details on the calibration procedures are given in \S \ref{section:wavcalib} and following sections.  The observing conditions over the four half nights nights were very good, with seeing estimates in the range 0.7\,-\,1.2 for targets observed at airmasses $<$\,$1.5$. Our targets are listed \hbox{in Table \ref{tab:targets}}.

\subsection{Proxima Centauri Observations}
\protect\label{section:obs_proxcen}
Proxima Centauri has been shown by \cite{endl08proxcen} to be stable to 3.11 \ms\ over a 7 year period and thus we consider this to be a good target to pursue as a calibrator for our techniques. Data taken with {\sc uves}, spanning five nights, with observations made on three nights and single night gaps, were obtained from the {\sc eso} data archive. These data were initially taken as part of a multi-wavelength survey of Proxima Centauri (GJ 551) and are presented in \cite{fuhrmeister11proxcen}. { Approximately} 560 spectra of Proxima Centauri were continuously recorded on each of the three nights on 2009 March 10, 12 \& 14, spanning 8 hours per night with altitudes corresponding to an airmass range of 2.41\,-\,1.27. A slit width of 1\arcsec\ gives a spectral resolution, R $\gtrsim$ 43,000 in the red arm of {\sc uves}, while the extreme airmass range of the observations led to seeing that varied from 1.4\arcsec\ at high airmass, down to $\sim$\,0.6\arcsec\ at low airmass. The CCD readout was binned in the wavelength direction by a factor of 2, resulting in an average pixel increment of 2.4 \kms. The rotation velocity of Proxima Centauri, at \hbox{\vsini\ = 2 \kms}, means that spectral resolution (equivalent to $\sim$\,6\,\kms) dominates the line width.

\subsection{Data extraction}
The data sets for both our ROPS sample (\S \ref{section:obs_rops}) and Proxima Centauri (\S \ref{section:obs_proxcen}) were flat field corrected by using combined exposures taken with an internal tungsten reference lamp. Since few counts are recorded in the reddest orders of the \mitll\ CCD (owing to the spectrograph efficiency and low quantum efficiency of the CCD longward of 1.0 \micron), where of order 10,000 counts could be achieved with 14 sec exposures compared with a peak of 40,000 counts, an additional 30 flatfield frames were taken in addition to the standard calibrations for the ROPS (\S \ref{section:obs_rops}) data set. The worst cosmic ray events were removed at the pre-extraction stage using the Starlink {\sc figaro} \citep{shortridge93figaro} routine {\sc bclean} (The Starlink software is currently distributed by the Joint Astronomy Centre\footnote{http://starlink.jach.hawaii.edu/starlink}). The spectra were extracted using {\sc echomop}'s implementation of the optimal extraction algorithm developed by \citet{horne86extopt}. {\sc echomop} rejects all but the strongest sky lines \citep{barnes07b} and propagates error information based on photon statistics and readout noise throughout the extraction process. 

\begin{figure}
\begin{center}
\includegraphics[width=6.5cm,height=8.5cm,angle=270]{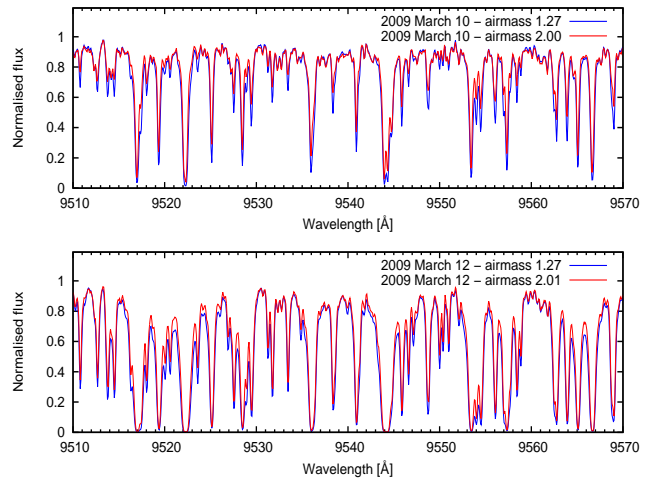} \\
\end{center}   
\caption{Spectral region 9510\,-\,9570\,\AA\ illustrating the change in humidity between 2009 March 10 (low humidity: 2\,-\,14 per cent) and 2009 March 12 (high humidity: 48\,-\,54 per cent) at Paranal. Note that some lines become strongly saturated when the humidity levels are high. Even those lines that do not saturate are highly variable in strength, with some lines (e.g. 9550\,-\,9551\,\AA) almost disappearing.}
\protect\label{fig:telluric_variation}
\end{figure}

\section{Wavelength calibration}
\protect\label{section:wavcalib}

Wavelength calibration at the \ms\ level is required if precision radial velocities are to be achieved. To this end, a great deal of effort has been expended in order to obtain accurate wavelengths for spectral calibration references (e.g. \citealt{gerstenkorn78}). Despite recent work that has identified new sources for calibration, suitable reference lines are often limited in the wavelength regions that they span. \cite{mahadevan09gascells} have identified a number of molecular gas cells that could be used to span the H band, while {\sc LASER} comb technology has also been used to demonstrate \hbox{$\sim$\,10 \ms}\ precision on sky in the H band \citep{ycas12lasercomb}. Although new calibration sources, rich in lines, have also been identified in the red part of the optical \citep{redman11uranium}, ThAr still remains the most regularly used and only available calibration source for optical and infrared high resolution spectrometers, although with relatively few lines in the near infrared ($>$\,$1$\micron).

\begin{table*}
\begin{tabular}{lccccccccccc}
\hline
Star     		  & SpT   & Imag & Exp           & $v$ sin $i$ & S/N          & Mean S/N      & Nobs         & r.m.s.   & r.m.s.     & r.m.s.    & r.m.s.    \\
		          &       &      & [s]           & [kms$^{-1}$]&              &               &              &[ms$^{-1}$]&[ms$^{-1}$]&[ms$^{-1}$]&[ms$^{-1}$]\\    
                          &       &      &               &             & Extracted  &   Decon     &              & No corr  & L corr     & T corr    & L-T corr  \\
\hline
GJ 3076			  & M5V   & 10.9 & 400           &   17.1*     &  77 $\pm$ 10 & 6120          & 4            &  100.7   &    92.3    &    67.6  &    44.5   \\ 
GJ 1002   		  & M5.5V & 10.2 & 300           &   $\leq$3   & 106 $\pm$ 12 & 9110          & 4            &   29.4   &     5.1    &    12.9  &    23.6   \\ 
GJ 1061   		  & M5.5V & 9.5  & 300           &   $\leq$5   & 142 $\pm$ 12 & 12150         & 5            &   4.23   &     2.4    &     2.4  &     2.8   \\ 
LP 759-25		  & M5.5V & 13.7 & 1500          &   13        &  38 $\pm$  5 & 2810          & 4            &  106.8   &    79.9    &    70.6  &    65.9   \\ 
GJ 3146			  & M5.5V & 11.3 & 600           &   12.4*     &  60 $\pm$ 10 & 4920          & 4            &   87.2   &    47.1    &    80.2  &    7.75   \\ 
GJ 3128   		  & M6V   & 11.1 & 350           &   $\leq$5   &  65 $\pm$  3 & 5590          & 4            &   24.4   &    11.5    &    24.1  &    15.6   \\ 
Proxima Centauri	  & M6V   & 6.9  & 100$^\dagger$ &   2         & 191 $\pm$ 22 & 12900         & 561	     &    5.2   &     -      &      -   &      -    \\
GJ 4281			  & M6.5V & 12.7 & 1200          &   7         &  49 $\pm$  6 & 4240          & 4            &   36.7   &    11.7    &    12.0  &    15.3   \\ 
SO J025300.5+165258	  & M7V	  & 10.7 & 350           &   $\leq$5   &  95 $\pm$ 12 & 8280          & 4            &   15.2   &    12.4    &    12.5  &    14.6   \\ 
LP 888-18		  & M7.5V & 13.7 & 1500          &   $\leq$3   &  36 $\pm$  3 & 2790          & 4            &   45.3   &    35.3    &    31.6  &    38.0   \\ 
LHS 132			  & M8V   & 13.8 & 500           &   $\leq$5   &  37 $\pm$  2 & 3160          & 4            &   12.3   &    12.3    &     7.7  &     9.09  \\ 
2MASS J23062928-0502285   & M8V   & 14.0 & 1500          &   6         &  38 $\pm$  1 & 2890          & 4            &   29.0   &    14.2    &    16.9  &    10.0   \\ 
LHS 1367		  & M8V   & 13.9 & 1500          &   $\leq$5   &  32 $\pm$  3 & 2470          & 4            &   22.7   &    15.3    &    16.1  &    22.5   \\ 
LP 412-31		  & M8V   & 14   & 1200          &   12        &  26 $\pm$  7 & 1850          & 3            &  253.2   &   222.6    &   248.9  &   119.6   \\ 
2MASS J23312174-2749500	  & M8.5V & 14.0 & 1500          &   6         &  37 $\pm$  2 & 2890          & 4            &   37.2   &    36.7    &    29.5  &    22.3   \\ 
2MASS J03341218-4953322	  & M9V   & 14.1 & 1500          &   8         &  33 $\pm$  2 & 2810          & 4            &   11.2   &     6.37   &     6.92 &     8.37  \\ 

\hline
\end{tabular}
\vskip 2mm
\caption{List of targets observed with {\sc uves} with estimated spectral types, I band magnitudes, { exposure times and \vsini\ values (columns 1 to 5)}. The measured \vsini\ values are taken from Mohanty \& Basri (2003), Jenkins et al. (2009) and Reiners \& Basri (2010). We derived \vsini s for GJ 3076 and GJ 3146 (denoted by a *) using the procedures we adopted in Jenkins et al. (2009). We also list details for Proxima Centauri { and the mean r.m.s. of 5.2 \ms, after atmospheric correction for all three nights, is given}. The exposure times$^\dagger$ for Proxima Centauri were variable, ranging between 11 secs and 500 secs, however 74 per cent of the observations were made with 100 sec exposures. Extracted S/N ratio and S/N ratio after deconvolution are tabulated in columns 6 \& 7. { Column 8 lists the total number of observations, $N_{obs}$ on each target and column 9 gives the r.m.s. scatter using an $N_{obs} - 1$ correction to account for the small number of observations for each object (see section \S \ref{section:results}). In columns 10, 11 \& 12, we list r.m.s. values after applying bisector corrections derived from the stellar line (L), telluric line (T), and both lines (L-T). Discussion of the results is given in \S \ref{section:results}.}}
\protect\label{tab:targets}
\end{table*}

\subsection{Master wavelength calibration}
\protect\label{section:wavmaster}

ThAr wavelengths published by \cite{lovis07thar} were used to identify stable lines for wavelength calibration. This line list is estimated to enable a calibration {(i.e. global)} r.m.s to better than { 20 \cms}\ for {\sc harps}. Pixel positions were initially identified for a single arc using a simple Gaussian fit. For each subsequent arc, a cross-match was made, followed by a multiple-Gaussian (up to three profiles) fit around each identified line using a Levenberg-Marquardt fitting algorithm \citep{press86} to obtain the pixel position of each line centre. The Lovis \& Pepe line list was optimised for {\sc harps} at R = 110,000, while our observations were made at R $\sim$ 50,000 necessitating rejection of some lines that showed blending. Using a multiple Gaussian fit enables the effect of any nearby lines to be accounted for in the fit that also included a first order (straight line) background. Any lines closer than the instrumental FWHM were not used. Finally for each order, any remaining outliers were removed after fitting a cubic-polynomial. In addition, any lines that were not consistently yielding a good fit for all arc frames throughout both nights (to within 3-sigma of the cubic fits) were removed. 

The ThAr observation following each star on the second night was chosen arbitrarily as the reference solution for that star. The wavelengths were then incrementally updated for all other observations of each star using the methods that we describe in \S \ref{section:telluric_rv_procedure} and \S \ref{section:rops_thar}, which are aimed at minimising systematics in the wavelength solutions from one observation to the next. A total wavelength span of 6519 \AA\ to 10252 \AA\ is covered by the \eev\ and \mitll\ chips at the non standard 840 nm setting of {\sc uves}. An order that falls between the two CCDs can not be used and must be accounted for correctly in the two dimensional solution. The candidate ThAr lines were subjected to a two dimensional fit of wavelength vs extracted order (cross-dispersion) for each CCD independently. For each star, an arbitrary reference solution with a two dimensional polynomial fit using 4 coefficients in the wavelength direction and 6 coefficients in the cross-dispersion (order) direction was made:

\begin{equation}
 \lambda(x,y) = \sum_{i=0}^3 a_{i} x^{i} \sum_{j=0}^{5} b_{j} y^{j}
\end{equation}

where $a$ and $b$ are the polynomial coefficients that we fit for. $x$ and $y$ are the pixel number and order number respectively and $i$ and $j$ are the powers in $x$ and $y$ for each coefficient. By iteratively rejecting outlying pixels from the fit, we found that of the input 573 lines, clipping the furthest outliers yielded the most consistent fit from one solution to the next. Typically 15-20 lines were rejected before a final fit was produced for each observation. The zero point r.m.s. (i.e. the r.m.s. by combining all lines) for the {\em master wavelength calibrations} is found to be \hbox{$\sim$\,$5$\,-\,$5.5$ \ms}\ and \hbox{$\sim$\,$6$\,-\,$6.5$ \ms}\ for the \eev\ and \mitll\ chips respectively and represents the goodness of fit of the polynomial. { These values are dominated by a systematic difference between the wavelengths and the two dimensional fit. The {\em variability in the wavelength solution} for a given set of radial velocity measurements (i.e. for each star) is thus important and ultimately determines the precision that can be achieved. We discuss this further \S \ref{section:rops_thar}, but note here that this variability is an order of magnitude smaller (i.e. $<$\,1\,\ms) than the zero point r.m.s. values quoted above.}

The appropriate wavelength solution for each observation can obtained through a {\em simultaneous} measurement, by using the telluric lines, or a {\em near-simultaneous} measurement by using the nearby ThAr reference frame. In each instance, the corrections are determined as pixel shifts and applied to the master wavelength solution. This procedure enables wavelength corrections to be applied, allowing for low order shifts and stretches (due to mechanical effects and temperature/pressure changes). In other words, allowing more degrees of freedom for each solution can lead to poor fits in the first and last order, near the order edges, and in regions where there may be fewer lines. Low order corrections correctly describe the changes in the instrument while minimising variability in the fits. We describe the two methods adopted in this paper for updating the wavelength, using telluric lines (\S \ref{section:telluric_rv_procedure}) and ThAr lines (\S \ref{section:rops_thar}).

\section{Tellurics as a wavelength reference}
\protect\label{section:tellurics}

The benefit of utilising telluric lines to obtain a local wavelength solution is that the wavelengths are derived from the very observation of the star itself and are therefore simultaneous. The telluric spectrum essentially follows the same light path as the star through the earth's atmosphere, the telescope and the spectrograph, and is thus subject to the same systematics. The calibration procedure could be seen as analogous to that first adopted by \cite{marcy92iodine} and \cite{butler96iodine} if the atmosphere of the Earth could be well characterised and calibrated for. In addition, at the time of observations, there were no optical 8 m class spectrometers that enable simultaneous ThAr observations to be made. 

The stability of telluric lines as reference fiducials has been investigated by a number of authors. \cite{griffin73}, for example, made some initial attempts to identify lines in the 6841\,-\,7424\,\AA\ region, estimating uncertainties at the 1-2 m\AA\ level, equivalent to $\sim$\,40\,-\,90 \ms. The most complete list of ab initio line strengths and positions for water have now been calculated by \cite{barber06water} and are now routinely used in model atmosphere databases that supply molecular information for many molecules \cite{rothman09hitran}. Bands of telluric molecular absorption lines pose a challenge for any ground based observations and are seen from the mid-optical, becoming stronger and wider into the infra-red. In the red-optical, at wavelengths greater than 6500 \AA, significant O$_2$ absorption bands, with bandheads at \hbox{$\sim6865$}\,\AA\ and \hbox{$\sim7595$}\,\AA\ appear, the latter showing strong absorption with saturation in some lines. H$_2$O bandheads at 6450\,\AA, 7170\,\AA, 8100\,\AA\ exhibit increasing widths from \hbox{$\sim$\,$100$}\,\AA\ to several $100$\,\AA, however the band covering 8890\,\AA\ to 9950\,\AA\ is by far the most extensive at wavelengths short of 1\micron. \cite{gray06tellurics} were able to achieve empirical precisions of $\sim$\,$25$\,\ms\ using strong H$_2$O absorption lines in the 6222\,-\,6254\,\AA\ region formed in the optical path of the Coud\'{e} \'{e}chelle spectrograph used. This procedure had the advantage of minimising atmospheric projection effects such as change in airmass. 

\cite{figueira10stability} instead took advantage of night long observations made on bright stable stars with the {\sc harps}, located at the ESO 3.6 m at La Silla. They found clear nightly trends of the radial velocity variations of the O$_2$ absorption band as measured in the spectra of $\tau$ Ceti (HD 10700), $\mu$ Ara (HD 160691) and $\epsilon$ Eri (HD 20794). Empirical fits were made to the velocities using a simple model that included a linear airmass term (fixed, with magnitude, of $\sim$ $20$\,ms$^{-1}$), a projection of the wind velocity along the line of sight of the telescope (encompassing both magnitude and direction), and a fixed calibration offset term. Such a procedure enabled typical measurement precisions over week long timescales of 4.5\,-10\,\ms\ to be made for observations at less than 1.5 airmass ($>$\,$41.8$\degs) and $2.4$\,-\,$4$\,\ms\ when restricting observations to less than 1.1 airmass ($>$\,$65.4$\degs). Over a period of 6 years, the precision { was found to be} of order 10 \ms.

\begin{figure*}
\begin{center}
\includegraphics[width=8cm,height=17.5cm,angle=270]{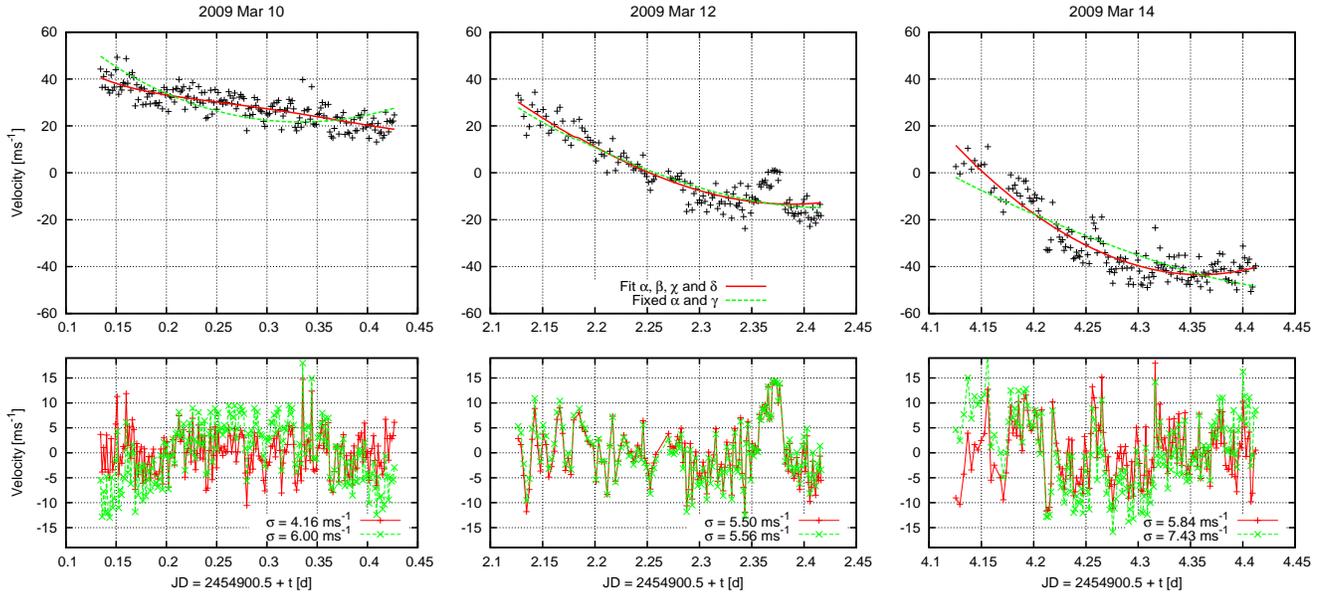} \\
\end{center}   
\caption{Radial velocities of Proxima Centauri for observations made on 2009 March 10, 12 \& 14. The top panels show the heliocentrically corrected radial velocities on each night along with the fits which account for changes in airmass, wind velocity, direction and offset. { The solid/red lines indicate the fits made to each night individually while the green/dashed lines are for fits made with fixed $\gamma = 17.75$}. Seeing effects in the 1\arcsec\ slit, which varied between a maximum of 1.4\arcsec\ at high airmass to a minimum of 0.6\arcsec\ at low airmass were not accounted for during the modelling. The residuals are plotted in the bottom panels { (line type and colour corresponds to the top panels). The nightly-subtracted residuals vary between 4.16 \ms\ and 5.84 \ms, while the corresponding $\gamma = 17.75$ fit residuals give r.m.s. values of 5.56\,-\,7.43 \ms.}}
\protect\label{fig:proxcen_rvs}
\end{figure*} 

\subsection{Precision radial velocities of Proxima Centauri}
\protect\label{section:proxcen_rvs}

The opportunity to study the stability of telluric lines alone for updating the wavelength solution and providing a stable cross-correlation reference against which to make precision radial velocity measurements is afforded by the archival observations of Proxima Centauri,  already outlined in \S \ref{section:obs_proxcen} and initially published in \cite{fuhrmeister11proxcen}. We intended to characterise the stability and behaviour of {\sc uves} for our radial velocity measurement technique by making use of the $\sim$\,560 { archival}  observations { taken} over three nights, with an intention of extending the method to our ROPS sample. In addition, this kind of study is not possible with our 2012 July observations since we only observed each target once per night, which precludes monitoring stability on minute to hour-long timescales. \cite{kurster99proxcen} showed that Proxima Centauri is stable to the \hbox{54 \ms}\ level, while more recent results from \cite{endl08proxcen} have shown it to be stable to \hbox{3.11 \ms}\ over a \hbox{7 year} period, but quote an average propagated uncertainty of \hbox{2.34 \ms}\ in their measurements, indicating an additional unaccounted for source of noise.


The seeing variations of 1.4\arcsec\ at high airmass, down to $\sim$\,0.6\arcsec\ at low airmass, when viewed with a 1\arcsec\ slit offer a less than ideal match since the star does not completely fill the slit. This results in changes in illumination of the \'{e}chelle, leading to radial velocities that can potentially vary at the \ms\ to several tens of \ms\ level. 
Since the CCD readout was binned by a factor of two in the wavelength direction, the mean pixel increment of 2.4 \kms\ is twice that of the full 1.2 \kms\ mean readout increment used for the ROPS targets. {Only one ThAr frame per night} was recorded during the automated calibration procedures executed by {\sc uves} each night. As a result, it is impossible to track any drift of the spectrograph through the night, or in this instance, to investigate our ability to use the ThAr frames as a near-simultaneous reference fiducial. During extraction, we also discovered that on { 2009}\ March 10 and 14, a regular half hour, cyclic shift of order 1 \kms\ appears in the radial velocities. The origin of this cyclical behaviour is unclear, but it appears to coincide with times at which the seeing was very good. We believe that it is related to the mismatch of seeing and slit width where the autoguider may have been fooled into making only occasional corrections that have resulted in significant \'{e}chelle illumination change.

\subsection{Radial velocity measurement procedure}
\protect\label{section:telluric_rv_procedure}

Starting with the two dimensional wavelength solution described in \S \ref{section:wavmaster}, a method of updating the local wavelength solution for each observation must be obtained. No special wavelength calibrations were made during the observing sequence of Proxima Centauri, { and} only one { ThAr spectrum} was recorded during the standard calibrations { for each night}. We therefore investigated the use of the abundant H$_2$O and O$_2$ in the red-optical to update the wavelength solutions.

The weather conditions on the first night were particularly dry, with relative humidity variations in the \hbox{2\,-\,14 per cent} range (as recorded for the telescope dome in the observation headers). On the second and third nights, the relative humidity varied in the ranges \hbox{48\,-\,54 \%} and \hbox{22\,-\,28 per cent}. The increased water column is clearly evident in the H$_2$O lines as illustrated in Fig. \ref{fig:telluric_variation}. This additionally serves to illustrate why the use of water lines for precision radial velocity work can prove challenging. With careful selection, it is in fact possible to select H$_2$O lines that are not blended with other lines and that also do not vary so greatly in strength as to become insignificant relative to the continuum level noise. Since Fig. \ref{fig:telluric_variation} illustrates the extremes of the telluric line variations during the Proxima Centauri observations, we found that the optimal procedure was to manually select the appropriate lines that fit these criteria. Over the 0.65\,-\,1.025 \micron\ interval, an initial list of tellurics comprising $\sim$ 1700 lines, with normalised line depths in the range 0.1\,-\,1.0, results in a subset of only $\sim$\,$300$ non-blended H$_2$O and O$_2$ lines with normalised lines in the 0.6\,-\,0.95 range. Changes in instrumental resolution are likely to affect the selection, with the expectation that more lines could be used with a higher instrumental resolution.

Despite selecting only the strongest unblended H$_2$O lines, we found that the most stable procedure entailed utilising {\em only} the O$_2$ lines that are recorded in two bands on the \eev\ chip. The use of O$_2$ lines was advocated and adopted by \cite{figueira10stability} since they are more stable than H$_2$O lines which occur in a very narrow layer and are highly variable, being correlated with weather and humidity patterns. We thus made use of only the orders recorded on this chip for the Proxima Centauri data set, which span 6519\,-\,8313\,AA. Since the two O$_2$ bands span 5 orders in total, with some lines recorded twice, we make use of the full information by determining the shift of every recorded instance of each line. A mask is made, to include all the O$_2$ lines within to 4 FWHM. Only these lines are used to determine the transform.

We found the most reliable procedure for updating the wavelength solutions via telluric lines is to calculate the transform that maps the reference spectrum to each individual observation in turn.  The normalised master spectrum $t_j$ is thus scaled to the current normalised observed spectrum, $s_j$ by minimising the function
\begin{equation}
\chi^2 = \sum_{i} \left( \frac{s_i - (\xi_0 + \xi_1t_i + \xi_2t_i^2)}{ \sigma_i + \tau_i} \right)^2
\end{equation}
\noindent
where 
\begin{equation}
f_i = \xi_0 + \xi_1t_i + \xi_2t_i^2
\end{equation}
\noindent
is the transformed normalised master spectrum and $\xi_1$, $\xi_2$ \& $\xi_3$ are the quadratic transform coefficients for each O$_2$ line \hbox{pixel, $i$,} designated by the mask. $\sigma_i$ and $\tau_i$ are the uncertainties on the observed spectrum and the master spectrum respectively. This procedure is implemented such that all mask designated lines are fitted simultaneously. In other words, the same transform can be applied to all orders to update the wavelength solution. Since $\xi_1$, $\xi_2$ \& $\xi_3$ are in pixel units, the wavelength increment per pixel is calculated from the master wavelength frame for all pixels over all the orders used for determining radial velocities. The master wavelength increment map is multiplied by the pixel increments and added to the master wavelengths to update the wavelength solution. 

{ As in \cite{barnes12rops} we carry out a least squares deconvolution using line lists that represent both the telluric line and the stellar line positions}. We use the Line By Line Radiative Transfer Model ({\sc lblrtm}) code \citep{clough92,clough05} to obtain telluric line lists, while we derived the stellar line lists empirically. In the latter case, we used high S/N observations of \hbox{GJ 1061} { made with a 0.4\arcsec\ slit}. The \hbox{GJ 1061} line list was used for deconvolution of the M5V\,-\,M7V targets. For the M7.5V\,-\,M9V targets we used the spectra of \hbox{LHS 132}\ (aligned and co-added to augment the S/N ratio). { The procedure for derivation of the stellar templates used in this paper is given in Appendix \ref{section:appendix_rvs}}. Two high S/N ratio lines are thus calculated for each spectrum, with the final velocity calculation being made by subtracting the telluric line position from the stellar line position (measured via cross-correlation).

\subsection{Radial velocity stability of Proxima Centauri}
\protect\label{section:proxcen_stability}

The radial velocities for the three nights are shown in Fig. \ref{fig:proxcen_rvs}. { All radial velocities presented in this section are listed in \ref{tab:appendix_table_proxcen}.} There is a clear trend during each night and an offset, particularly when the first night is compared with the second and third nights. Both the slope, curvature and offset changes from night to night. We have used the empirical procedure outlined in \cite{figueira10stability} to model the trends seen in the RVs on each night. The radial velocity correction

\begin{equation}
 \Omega = \alpha \left(\frac{1}{sin(\theta)} - 1 \right) + \beta cos(\theta) cos(\phi - \delta ) + \gamma
 \protect\label{equation_omega}
\end{equation}

was shown to be sufficient to adequately remove atmospheric effects. The parameters $\alpha$, $\beta$, $\gamma$ and $\delta$ can be determined when observations are made throughout the night at the telescope elevations ($\theta$) and azimuth angles ($\phi$) of a fixed target. $\alpha$ represents the linear radial velocity drift per airmass ($1/sin(\theta)$) due to changes in the line shape as different layers of the atmosphere are sampled. $\beta$ is effectively the wind speed at the time of the observation and $\delta$ is the wind direction. $\gamma$ is an additional offset term that describes the offset of the observations from zero, when all other terms are zero; in our case, this the heliocentric velocity correction. We have enabled all parameters to be fit in order to optimise the fit for each individual night. { After subtracting the nightly fits, the residuals yield r.m.s. values of $\sigma = 4.16$,\ $5.50$\ \&\ $5.84$ \ms\ on each of 2009 March 10, 12 \& 14 respectively (Fig. \ref{fig:proxcen_rvs})}. { See also \hbox{Table \ref{tab:appendix_table_proxcen}} for a list of all corrected velocities (column 4, entitled ``I corr'').} These values appear reasonable considering the expected Poisson limited S/N of $\sim$2 \ms\ \citep{barnes13ropacs}. From { previously unpublished} archival {\sc harps} data\footnote{http\://archive.eso.org/eso/eso\_archive\_main.html} and {\sc uves} observations \citep{zechmeister09uves}, we find the radial velocity of Proxima Centauri to show r.m.s. scatter at the \hbox{2.3 \ms}\ (27 observations) and \hbox{4.3 \ms}\ (339 observations) levels respectively { (Tuomi et al. 2013, MNRAS, submitted)}.

The typical wind speed values we determine (130, 150 and 190 \ms\ for each night) are large and potentially not physically realistic. In addition, we find respective values for $\alpha$, the variation per airmass, of 31, 11 \& 23 \ms\ while the value of $\gamma$ varies between { -169.1 \ms\ and 100.9 \ms}\ (i.e. 270 \ms\ variation). As noted by 
\cite{figueira10stability}, $\gamma$ and $\alpha$ should be fixed. However the observations are not ideal, with varying humidity (see \cite{gray06tellurics} for a discussion of temperature, pressure and humidity effects, that can reach \kms\ levels). The additional problems with the cyclical behaviour during good seeing and the apparent trend of uncorrected radial velocity drift with seeing, especially when the seeing FWHM falls in the 0.6\,-\,0.8\,\arcsec\ range in the 1\,\arcsec\ slit, are likely to yield systematics. For this reason, we believe that the data are not able to reliably constrain wind speed values and directions for the Proxima Centauri observations, unlike the highly stabilised {\sc harps} observations of $\tau$ Ceti. Nevertheless, by holding $\gamma$ fixed at the { mean velocity (for the three nights)} and fixing the 17.75 \ms\ value for $\alpha$ found by \cite{figueira10stability}, the corresponding corrected radial velocity r.m.s. values for each night are $\sigma = 6.00$,\ $5.60$\ \&\ $7.43$ \ms\ on March 10, 12 \& 14 respectively. { The corrected velocities using this procedure are listed in \hbox{Table \ref{tab:appendix_table_proxcen}} (column 5, entitled ``A corr'').} More reasonable wind speeds of 115, 74 and 53 \ms\ are found, but again we stress that these are probably biased by the unconstrained effects discussed above. Most notably, the curvature is not fit well in these fits (Fig. \ref{fig:proxcen_rvs}, upper panel green curves) indicating the probable involvement of seeing variations. { For comparison, when considering the data taken with an airmass range up to 1.5, \cite{figueira10stability} found r.m.s. scatter of between \hbox{4.54 \ms}\ and \hbox{5.81 \ms}\ for \hbox{$\tau$ Ceti} (G8.5V) using the same method as described here. The radial velocity of $\tau$ Ceti is known to be very stable with a standard deviation of \hbox{1.7 \ms}\ \cite{pepe11}.}

\subsection{Concluding remarks}
\protect\label{section:concluding_proxcen}

The study in this section was motivated by a desire to characterise a {\em simultaneous} reference fiducial in order to obtain a local wavelength solution for our deconvolution procedure. With a few caveats, we are able to reproduce similar precision with an M6V star (Proxima Centauri) to that achieved with a G8V star ($\tau$ Ceti) with {\sc harps}. Undoubtedly, a stabilised spectrograph, a narrower slit (or at least a slit width well matched with the median seeing) should remove some of the additional trends in the data that equation \ref{equation_omega} cannot describe. Despite these promising findings, the major drawback of this procedure is that regular observations of a single target throughout each night would be necessary for successful implementation. We would never realistically expect to observe a given target at such a range of airmasses, and indeed \cite{figueira10stability} found that restricting observations to a narrower { airmass} range was necessary to achieve the precisions reported. 

Given that the trends throughout each night are also approximately linear or quadratic, correcting for atmospheric effects with a four parameter fit such as Equation \ref{equation_omega} clearly requires very high S/N ratio. Obtaining few \ms\ precision via this method has been possible for Proxima Centauri observations that enable S/N ratios of a few hundred. However, typical observations of late M stars will only achieve S/N ratios of several tens, which will more severely restrict the precision achievable. Internal calibration references are therefore always a preferred, and more realistic option for obtaining the local wavelength solution for deconvolution. We thus subsequently adopt this procedure for our ROPS sample of late M dwarfs, described in the following sections.


\begin{figure*}
\begin{center}
\includegraphics[width=80mm,height=125mm,angle=270]{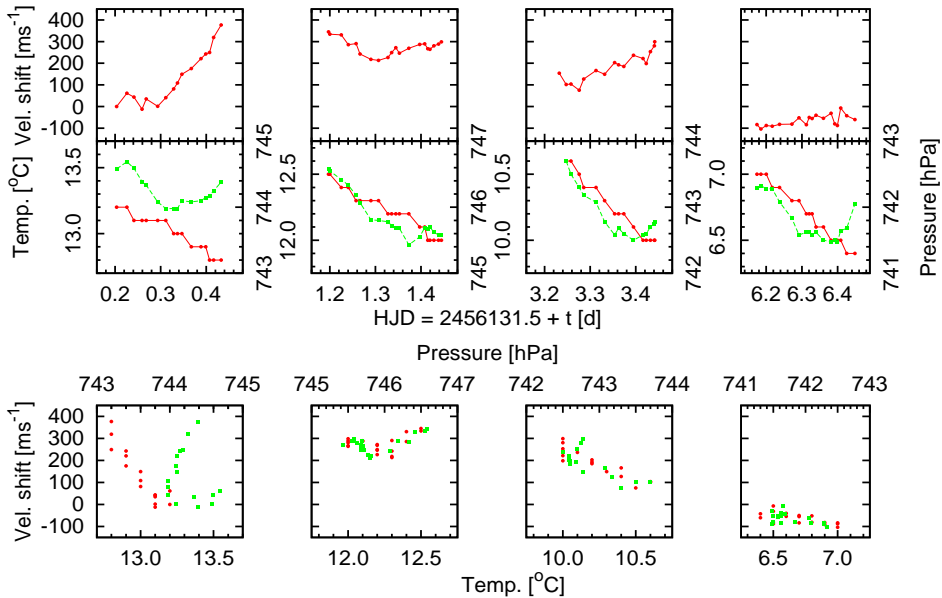} \\
\end{center}   
\caption{Stability of {\sc uves} during observations in July 2012. The top panels show the drift in \ms\ on each night and the middle panels plot the temperature of the red camera. While 0.4\degs\,-\,0.6\degs\ drift is seen on each night, the absolute temperature values are different. The bottom panels show the drift vs the temperature. Temperatures are plotted as filled red circles (scales on the left and bottom axes), while pressure is plotted as filled green squares (scale in hPa on the right and top axes).}
\protect\label{fig:stability}
\end{figure*} 

\section{{\sc uves} observations of a late M dwarf sample}
\protect\label{section:rops_obs}

For the late M stars observed with UVES, our strategy comprised of observing the same sequence of 15 targets during each of four half nights. The observations were made over a six day period on 2012 July 23, 24, 26 \& 29. This enables a time span that is sufficient to discern short period signals of order a few days. Since we are unable to implement the procedure described in the previous section, which made use of the telluric lines to update the wavelength solution for deconvolution of each spectrum (see \S \ref{section:concluding_proxcen}), we used the near-simultaneous ThAr frame recorded after each observation as a local wavelength solution.

\subsection{Radial velocity stability of {\sc uves}}
\protect\label{section:uves_stability}

In B12, we determined an incremental drift relative to a reference wavelength solution in order to obtain the local wavelength solution in each order. The {\sc mike} spectrograph however exhibited shifts of up to a few hundred \ms\ over short time scales, which we attributed to mechanical stability and possible gravitational settling of the dewar as the coolant boils off during the night. {\sc uves} appears to exhibit a much more predictable behaviour in that a more monotonic drift in wavelength is seen through a single night, although there is an offset between each night as shown in Fig. \ref{fig:stability} (top panels). Again the nightly offset may be related to both dewar refills and to re-configuration of {\sc uves} which regularly observes at different wavelengths. Shifts of order 50 \ms\ can be expected with {\sc uves} when different ThAr spectra are taken after changing the instrument configuration\footnote{http://www.eso.org/sci/facilities/paranal/instruments/uves/doc}. In addition, shifts of order $1/20$ pixel per \hbox{1 hPa} (millibar) change in pressure and the same shift for a change of 0.3\degs\ in temperature are typical. The recorded 0.4\degs\,-\,0.6\degs\ variation throughout each night (Fig. 1, filled red circles) during our observations, would thus lead us to expect 100\,-\,150 \ms\ { wavelength shift}. The pressure drift on each night is of order 1 hPa (Fig. 1, filled green squares) and hence presumably contributed to the observed drift. While attributing the observed shifts to temperature changes alone is in agreement with expectation on nights 2, 3 \& 4 of our observations, the first night, which was the least humid, showed \hbox{$\sim$\,$400$ \ms}\ drift through the night. At the same time, the temperatures were highest on the first night, possibly indicating that drift rate is correlated with temperature. This increased drift rate is discussed later in light of our derived radial velocities.

\subsection{Local ThAr wavelength solution}
\protect\label{section:rops_thar}

\begin{figure}
\begin{center}
\includegraphics[width=60mm,height=84mm,angle=270]{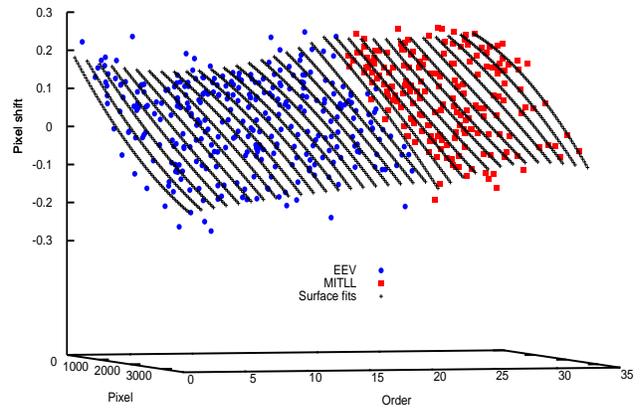} \\
\end{center}   
\caption{Example of the ThAr line pixel shifts for \eev\ (blue circles) and \mitll\ (red squares) CCDs for the 33 extracted orders. The black ``+" symbols represent the fitted 3 (wavelength) by 2 (cross-dispersion/order) polynomial surface. Shifts are relative to the master wavelength frame taken with each observation on the second night of observations.}
\protect\label{fig:pixelsurface}
\end{figure} 

Subsequent to obtaining a master solution for each star, as outlined in \S \ref{section:wavmaster}, we have adopted a method for obtaining the local wavelength solution for each frame that is different from that described in \S \ref{section:telluric_rv_procedure}, which made use of telluric lines. For our ROPS targets, we obtain the local wavelength frame taken after each observation by instead updating the wavelength positions of all the ThAr lines used to determine the master solution. The pixel positions of all the lines are calculated as outlined in \S \ref{section:wavcalib} before subtracting the line positions of the master wavelength frame. This procedure has the advantage that lower order corrections can then be applied to update the master wavelength solution. A two dimensional fit is made for pixel position vs order for all the measured pixels. In other words a two dimensional pixel shift surface is determined and we find that a polynomial of degree 3 (quadratic) in the wavelength direction and 2 (linear) in the cross-dispersion direction (Fig. \ref{fig:pixelsurface}) is sufficient to describe the drifting wavelength solution relative to the master solution which was calculated via a $4\times6$ polynomial (\S \ref{section:wavmaster}). The fitted pixel shift surface can be written as

\begin{equation}
 \Delta p(x,y) = \sum_{i=0}^2 a_{i} x^{i} \sum_{j=0}^{1} b_{j} y^{j}
\end{equation}

\noindent
where $\Delta p(x,y)$ is the pixel drift surface defined at each pixel, $x$, and extracted order number, $y$. The coefficients $a$ and $b$ scale the $x$ and $y$ terms of power $i$ and $j$ respectively. The pixel surfaces are converted to an updated wavelength surface by calculating wavelength increments from the master wavelength frame and adding to the master wavelength frame. This procedure has the advantage of maintaining stability as any order edge effects are minimised in a low-order fit. Zero point r.m.s. values in the wavelength solutions of \hbox{2.01 $\pm$ 0.20 \ms}\ and \hbox{2.63 $\pm$ 0.24 \ms} for the \eev\ and \mitll\ chips respectively are found. As already noted in \S \ref{section:wavcalib}, these values could be reduced by using additional calibration lamps, but we note that the 1-$\sigma$ variability is an order of magnitude lower at \hbox{20 c\ms\ and 24 c\ms}, and well below the photon noise precision that can be achieved with UVES using the techniques described in this paper.

\subsection{Radial velocities of 15 late M dwarfs}
\protect\label{section:results}



{ The mean-subtracted radial velocities for our ROPS targets are plotted in \hbox{Fig. \ref{fig:rvs}} with details of r.m.s. estimates listed in \hbox{Table \ref{tab:targets}}. Appendix \ref{section:appendix_rvs} gives full details of all radial velocities, which are listed in Tables \ref{tab:appendix_table_gj1061} \& \ref{tab:appendix_table_lhs132}}. The radial velocities are measured as outlined in B12 by subtracting the deconvolved telluric line position from the simultaneously observed stellar line. The line positions are measured by cross correlating each stellar line relative to the mean { deconvolved} stellar line for each target, and similarly for the telluric lines. We use the {\sc hcross} algorithm of \cite{heavens93redshifts} which is a modification of the \cite{tonry79} cross-correlation algorithm. {\sc hcross} utilises the theory of peaks in Gaussian noise to determine uncertainties in the cross-correlation peak. We have made a minor modification of the routine, which belongs to the Starlink package, {\sc figaro}, in order to directly output both the pixel shift, and shift uncertainty.

From Table \ref{tab:targets}, it can be seen that a range of exposure times and S/N values were obtained, depending on the brightness of the target, which ranged from $m_I$ = 9.5 to $m_I$ = 14.1. In addition, not all observed targets possess slow rotation, which we define as, at, or below the instrumental resolution of 54,000, or 5.55 \kms. \cite{jenkins09mdwarfs} found that at M6V, stars possess \hbox{\vsini\ = 8 \kms}\ on average, while this increases to \hbox{$\sim$\,15 \kms}\ by M9V. Table \ref{tab:targets} and Fig. \ref{fig:rvs} demonstrate that those stars { with} slower \vsini\ values on the whole appear to enable better radial velocity precision to be determined{ , as first noted by \cite{butler96iodine}. This is not surprising since the resolution is effectively degraded and line blending increases with increasing \vsini. The correlation between photon limited precision and r.m.s.\ {\em for a given \vsini} was also simulated in \cite{barnes12rops,barnes13ropacs}, and we further discuss and illustrate the ``excess'' r.m.s. (i.e. above that expected from \vsini\ and S/N ratio alone) in \S \ref{section:discussion}, \S \ref{section:halpha_proxy} and \hbox{Fig. \ref{fig:activity_vs_rms}}.}


{ For the early M dwarf sample targeted by {\sc harps}, \cite{bonfils13mdwarfs} found an anti-correlation when plotting bisector spans (BIS) against the measured radial velocities. For instance a clear correlation (with a Pearson's correlation of \hbox{$r$ = -0.81}) was identified for Gl 388 (AD Leo). Subtraction of the trend decreased the r.m.s. from 24 \ms\ to 14 \ms. We have calculated the BIS \citep{gray83,toner88,fiorenzano05bisectors} }for all our stars and subtracted the best fit linear trend with the derived RVs. The uncorrected RVs are listed in column 9 of Table \ref{tab:targets}, while the BIS corrected RVs are listed in column 10 and show that a number of our stars { also demonstrate trends that are} linked with the line bisector span (BIS). These stellar line BIS corrected velocities and subsequent r.m.s. values are plotted in Fig. \ref{fig:rvs}, { and we refer to these corrected values in the following discussion}. Significant improvements in the r.m.s are seen for a number of targets, where the r.m.s. is halved. The corrected RVs however show little improvement in the stars that exhibit the largest \vsini\ and derived r.m.s. values. Improvements are also seen if a correlation with the telluric BIS is removed (column 11), indicating that atmospheric variation may also contribute to limiting the precision that can be achieved using the methods outlined above. { Also, variability in the slit illumination (e.g. due to seeing changes) affects the instrumental point-spread-function, thus affecting both stellar and telluric lines to some degree. This will go some way to explaining why stellar or telluric lines can improve the measured r.m.s. However only the stellar lines contain line shape variability introduced by the star itself.} Finally, we have also investigated incorporating both the line and telluric BIS measurements. Since the final radial velocities are measured by subtracting the telluric line position from the stellar position, we also list RV-BIS corrections for a stellar-telluric BIS correction (column 12).

\begin{table*}
\begin{tabular}{lcccccccc}
\hline
\ \ \ \ \ Star     	& Sp Type & $v$ sin $i$ & Min     & Max       &      Min                            & Max                                 \\
                        &         & kms$^{-1}$  & EW (\AA)& EW (\AA)  & log$_{10}$(L$_{\rm H_{\alpha}}$/\,L$_{\rm bol}$)  & log$_{10}$(L$_{\rm H_{\alpha}}$/\,L$_{\rm bol}$)  \\
\hline
GJ 3076			& M5V     &   17.1      &   5.59  &     6.36  &    -3.81                            &  -3.76                              \\              
GJ 1002   		& M5.5V   &   $\leq$3   &   0.01  &     0.17  &    -6.51                            &  -5.42                              \\   
GJ 1061   		& M5.5V   &   $\leq$5   &   0.01  &     0.03  &    -6.74                            &  -6.12                              \\   
LP 759-25		& M5.5V   &   13        &   2.48  &     7.27  &    -4.08                            &  -3.79                              \\ 
GJ 3146			& M5.5V   &   12.4      &   2.80  &     3.92  &    -4.20                            &  -4.05                              \\ 
GJ 3128   		& M6V     &   $\leq$5   &   0.02  &     0.13  &    -6.36                            &  -5.62                              \\   
Proxima Centauri        & M6V     &   2         &   0.56  &     2.11  &    -5.00                            &  -4.43                              \\ 
GJ 4281			& M6.5V   &   7         &   0.85  &     1.06  &    -5.01                            &  -4.92                              \\  
SO J0253+1652	        & M7V	  &   $\leq$5   &   0.21  &     0.51  &    -5.61                            &  -5.58                              \\                 
LP 888-18		& M7.5V   &   $\leq$3   &   2.65  &     4.51  &    -4.83                            &  -4.60                              \\  
LHS 132			& M8V 	  &   $\leq$5   &   7.25  &    12.09  &    -4.36                            &  -4.14                              \\ 
2MJ2306-0502            & M8V 	  &   6         &   2.34  &     4.17  &    -4.85                            &  -4.60                              \\      
LHS 1367		& M8V     &   $\leq$5   &   2.59  &     6.11  &    -4.81                            &  -4.44                              \\
LP 412-31		& M8V     &   12        &  19.72  &    21.11  &    -3.93                            &  -3.90                              \\               
2MJ2331-27495	        & M8.5V   &   6         &   1.53  &     2.02  &    -5.12                            &  -5.01                              \\             
2MJ0334-49533	        & M9V     &   8         &   0.19  &     1.08  &    -6.14                            &  -5.39                              \\

\hline
\end{tabular}
\vskip 2mm
\caption{{ \ha\ variability for each object. Minimum and maximum \ha\ equivalent widths are listed for each object in columns 4 \& 5 respectively. The corresponding minimum and maximum log$_{10}$(L$_{\rm H_{\alpha}}$/\,L$_{\rm bol}$) are calculated from the appropriate models and listed in columns 6 \& 7 (see \S \ref{section:halpha_activity}).}}
\protect\label{tab:activity}
\end{table*}

\begin{figure*}
\begin{center}
\includegraphics[height=180mm,width=210mm,angle=270]{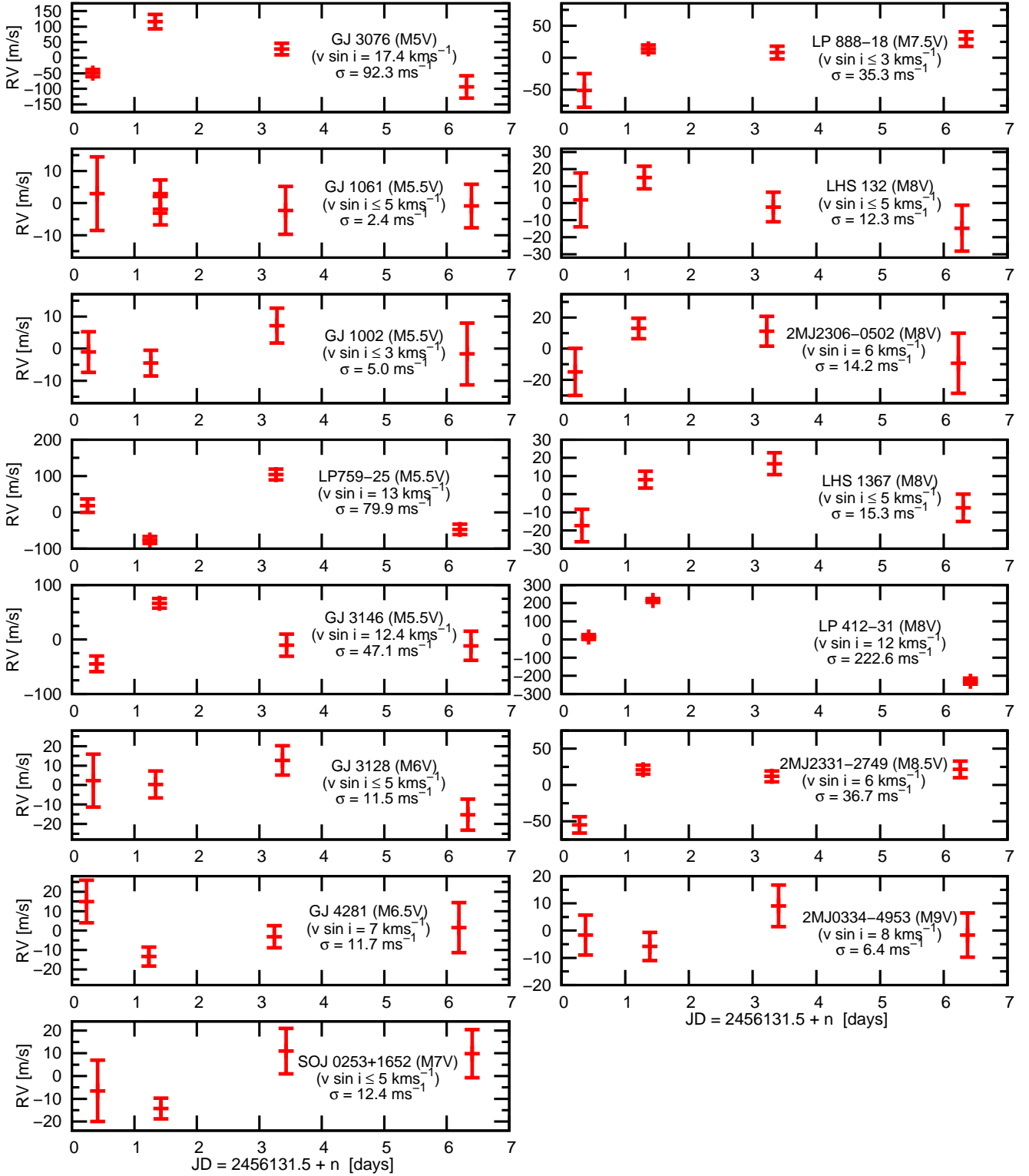} \\
\end{center}   
\caption{{ Heliocentrically corrected} radial velocities plotted for our 15 {\sc uves} ROPS targets. Observations were made on 2012 July 22, 23, 25 \& 28. The sample contains a total of seven M5V\,-\,M6.5V and eight M7V\,-\,M9V targets. A radial velocity precision of 2.4 \& 5.0 \ms\ is measured for quiet, slowly rotating targets at spectral type M5.5V (GJ 1061 and GJ 1002), while 6.4 \ms\ is found for our latest, M9V, target (2MASS J03341218-4953322). The targets showing higher r.m.s. in \hbox{Table 2} either exhibit significant rotation (\vsini\ $\gtrsim 10$ \kms), significant variability in the chromospheric indicators Ca {\sc ii} and \ha, or both.}
\protect\label{fig:rvs}
\end{figure*} 

\begin{figure*}
\begin{center}
\begin{tabular}{cc}
\includegraphics[height=8.4cm,angle=270]{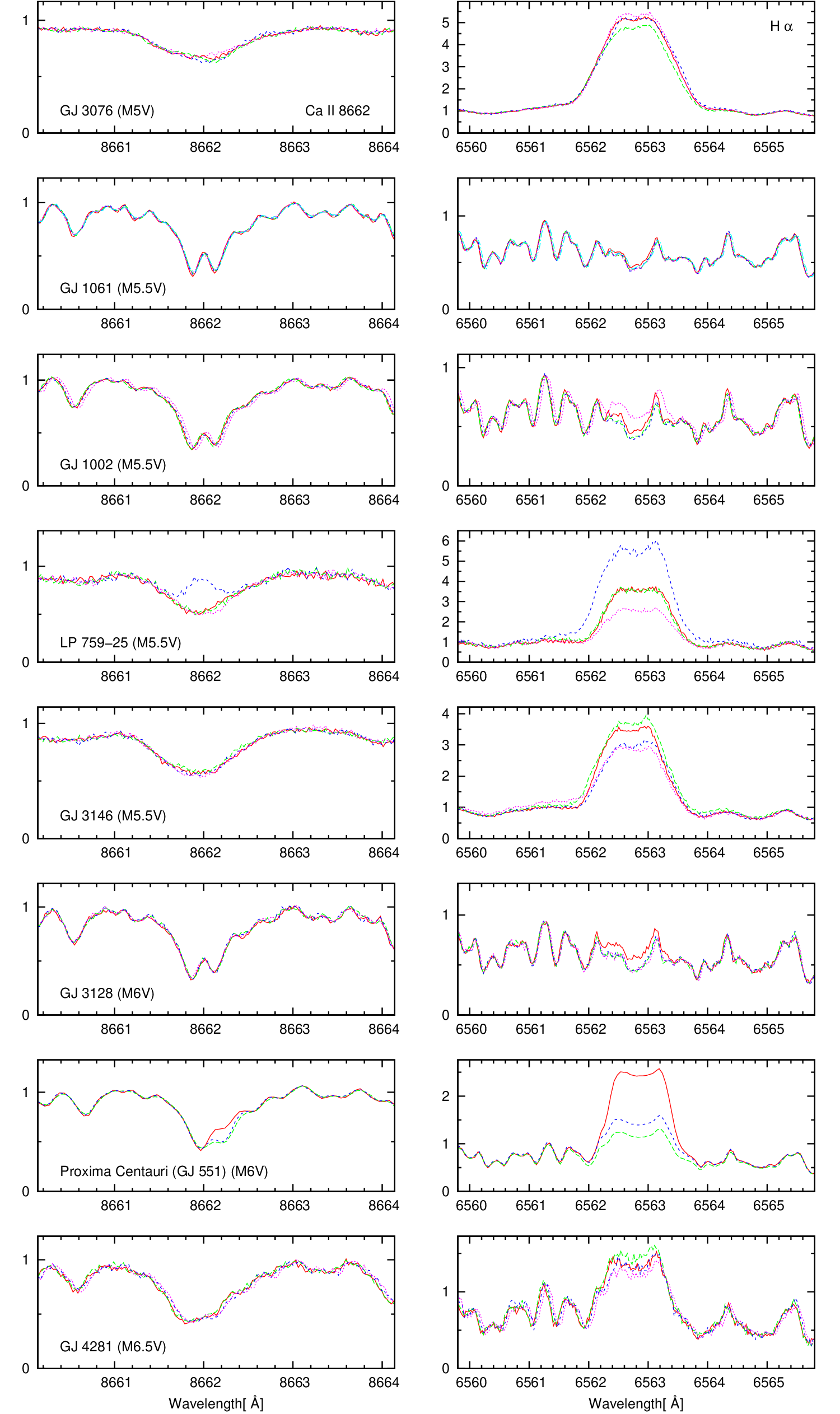} & \hspace{2mm}
\includegraphics[height=8.4cm,angle=270]{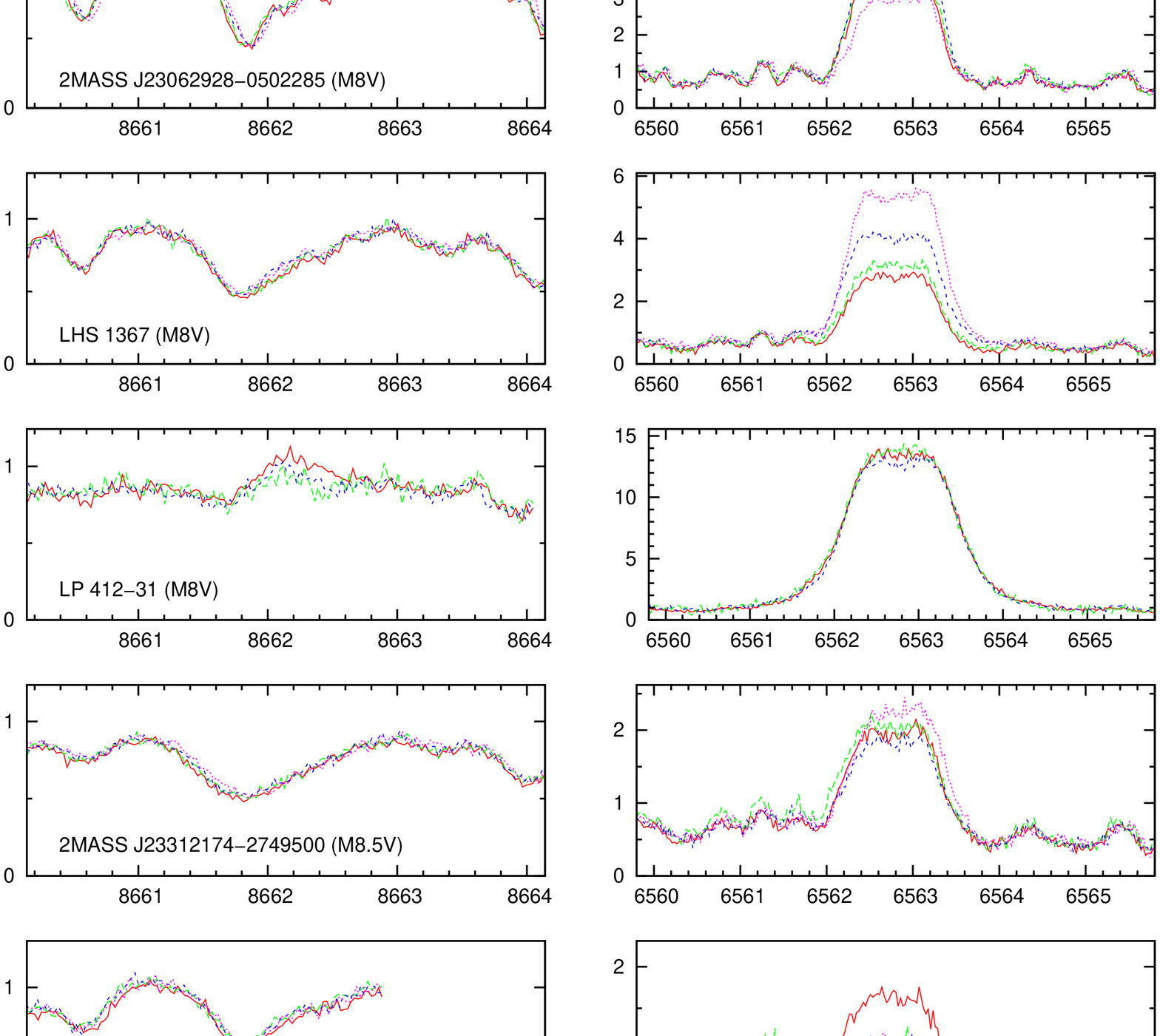} \\
\end{tabular}
\end{center}   
\caption{The Ca II 9662.14\,\AA\ profiles (left) and corresponding \ha\ (6552.80 \AA) lines (right) for all observations of the 15 ROPS targets listed in Table 1. For each target, the matched colour (online version) is used to signify that the Ca II 9662.14\,\AA\ and \ha\ lines were extracted from the same spectrum. We have also included profiles from the Proxima Centauri data set for the minimum, maximum and mean \ha\ emission and corresponding \hbox{Ca II 8662.14\,\AA}\ levels. The plotted wavelengths span 4 \AA\ for Ca II and 6 \AA\ for \ha. The blending of the Ca II 8662.14 \AA\ profile with the nearby Fe {\sc i} line at 8661.90 \AA\ is clearly seen in the slower rotating, earlier stars in the sample (e.g. GJ 1002, GJ 1061 \& GJ 3128). 2MASS J03341218-4953322 possesses a large radial velocity \hbox{(see Appendix A, $\Gamma_{\rm 2MJ03-49} = 73732.21$  \ms}),\ and since the Ca {\sc ii} line is located near the edge of the order, the spectrum appears truncated when the line is re-centred to 8662.14 \AA.}
\protect\label{fig:ca_ha}
\end{figure*} 

\subsection{Discussion}
\protect\label{section:discussion}

The r.m.s. velocities demonstrate that near-photon noise limited precision is achievable using our red optical survey. Following B12, where photon noise limited simulations { were} made with the {\sc mike} spectrograph at the 6.5m Magellan Clay telescope, we have estimated that 1.5\,-\,2 \ms\ should be achieved with {\sc uves} \citep{barnes13ropacs}. { The observations, in particular for \hbox{GJ 1061}, \hbox{GJ 1002} and 2MASS J03341218-4953322 (Table \ref{tab:targets}) thus show considerable improvements over recent measurements that have made use of telluric lines as a reference fiducial (e.g. \citealt{reiners09flare,rodler12keck,bailey12keck})}. The BIS corrected 2.4, 5.1 and 6.4 \ms\ measurements for these objects compare favourably with those that we obtained with {\sc harps} for the brightest targets in our sample. While GJ 1061 and GJ 1002 have not been actively monitored with {\sc harps}, 4 observations for each target { (that remain unpublished)} exist in the European Southern Observatory's archive. Using {\sc terra}, the Template-Enhanced Radial velocity Re-analysis Application \citep{anglada12terra}, a pipeline suite designed to improve the RVs achieved by the standard \harps\ Data Reduction Software (DRS), we have found \hbox{2.04 \ms}\ \& 2.32 \ms\ precisions for GJ 1061 and GJ 1002 { (see Table A4 for RVs).} We note that only the reddest orders of \harps\ in the very brightest mid M targets enable precision of a few \ms\ to be achieved.

Despite the sub-10 \ms\ r.m.s. values, a number of our stars exhibit radial velocities that are significantly in excess of the photon noise limited precision that we expect from our targets, { even when \vsini\ is taken into consideration. The RVs for the less stable targets indicate that rotation and activity may play a role in the observed larger r.m.s. values. As the average M6V star exhibits $v$\,sin\,$i$ =  8 kms$^{-1}$ \citep{jenkins09mdwarfs}, we expect velocity precisions of \hbox{$\sim$10 ms$^{-1}$} for \hbox{S/N = 30} \citep{barnes13ropacs}. However, while we predict photon noise limited precisions of \hbox{13, 15 and 20 \ms}\ for \hbox{GJ 3076}, LP 759-25 \& LP 412-31 respectively, they exhibit RVs that are an order of magnitude higher. The uncorrected RV values for these targets are also not significantly improved (at least relative to the photon-limited precision) when we include BIS corrections with the stellar lines or telluric lines. The best improvement is seen for the combined stellar and telluric line correction. While it is possible to select stellar lines for deconvolution that are free of any significant telluric lines (i.e. we { use} regions free of telluric lines with depths $>$\,$0.05$ of the normalised continuum), it is conversely not possible to select telluric line regions that are free of stellar lines. Any cross-contamination of the tellurics is thus more likely if the stellar lines show signs of activity variability. To ascertain whether the increased r.m.s. scatter may be { related to stellar variability}, { we investigate spectral lines that are sensitive to chromospheric activity in \S \ref{section:ca_ha}}

\subsubsection{The effect of instrumental drift on RV precision}
\protect\label{section:drift_effects}

The first night of our observations, 2012 July 23, was particularly dry and hence tellurics with smaller equivalent widths were derived, leading to RVs with larger error bars. Fig. \ref{fig:stability} also shows that the largest drift rates were observed with {\sc uves} on the first night.} Those targets that were observed during the { highest} rate of drift appear to show RV measurements with the greatest offset on each night. One might expect an improved velocity precision if each stellar observation were bracketed by ThAr observations, which would enable interpolation of the wavelength scale to the time centroid of the observation. { Applying this procedure} did not significantly improve our r.m.s. precision however, probably because the preceding ThAr was taken {\em before} the telescope was slewed to the new object. Unlike properly stabilised and fibre fed instruments {\sc uves} is located at one of the Nasmyth foci and is thus potentially subject to vibration and centripetal forces through slewing of the telescope from one target to the next. It is not clear whether movement of the telescope is able to affect the drift rate, but it doesn't necessarily appear to result in random changes in the drift {\em direction}. Bracketing every science exposure with { ThAr exposures (i.e. immediately before and after the observation)}, {\em with the telescope at fixed Right Ascension and Declination}, is like likely to enable further improvements in RV precision. This procedure will be adopted with any future observations.

\subsection{Chromospheric activity}
\protect\label{section:ca_ha}

{ The degree of stellar variability, as measured from chromospheric activity indicators in our target sample, varies considerably. While some of our more RV-stable targets such as \hbox{GJ 1061} and \hbox{GJ 1002} show low levels of chromospheric activity (e.g. flaring), others at similar spectral type and activity levels, such as Proxima Centauri, show higher levels of variability in lines such as \ha. The degree to which chromospheric activity significantly impacts upon measured radial velocities is not well known for mid to late M dwarfs. \cite{reiners09flare} found that the flaring activity, with 0.4 dex variability in \ha\ for the mid-M star, CN Leo, did not result in radial velocity deviations at the 10 \ms\ level, although a large flare event in that study did result in an RV deviation of several hundred \ms. The impact and correlation of activity variability with measured RVs in our ROPS sample is investigated in the following sections.}

{ 
\subsubsection{\ha\ as an activity indicator}
\protect\label{section:halpha_activity}

In order to monitor the chromospheric activity of each star (i.e. presence of active regions and flaring events), we have examined the \ha\ line, which is plotted for all observations in \hbox{Fig. \ref{fig:ca_ha}} (the \hbox{Ca {\sc ii}} 8662.14 \AA\ line, also plotted, is discussed in \S \ref{section:calcium_activity}). In the case of Proxima Centauri, we plot the minimum, mean and maximum \ha\ emission since there are a total of 561 observations in the 2009 data set.}


We have estimated the activity in our ROPS sample, by calculating \ha\ emission for all observations of each target. The \ha\ emission in each spectrum was calculated by measuring the equivalent width (EW) of the line. We adopted the procedure described in \cite{west04halpha}, by measuring the EW({H$_{\alpha}$}) relative to the normalised continuum. Following \cite{west08halpha}, the continuum regions are defined as 6555\,-\,6560 \AA\ and 6570\,-\,6575 \AA. Several of our targets, \hbox{GJ 1061}, \hbox{GJ 1002} and \hbox{GJ 3128}, have some or all measurements that yield negative EWs since the local continuum level is difficult to measure when \ha\ is barely visible. We have therefore assumed that all measurements are relative to the lowest measured EW which we assume is limited by the calculated EW uncertainty, as measured from the variances propagated during extraction. For any star with a significant emission EW, this uncertainty is negligible. Using flux calibrated spectra from nearby M stars, \cite{west08halpha} estimate $\chi$ values, the ratios of continuum flux around \ha\ to the bolometric flux. Using their tabulated values of $\chi$ for \ha\, we can determine \hbox{F$_{\rm H_{\alpha}}$/ F$_{\rm bol}$ = L$_{\rm H_{\alpha}}$/\,L$_{\rm bol}$ = $\chi({\rm {H_{\alpha}}})$ EW({H$_{\alpha}$})}. The same procedure was adopted by \cite{mohanty03activity} who instead of using flux calibrated observations, relied upon the models of \cite{allard01} to estimate $\chi$. { Luminosities are presented in the form}, log$_{10}$(L$_{\rm H_{\alpha}}$/\,L$_{\rm bol}$), which are given for each star in \hbox{Table \ref{tab:activity}}. It is immediately evident that the majority of stars show some degree of variability. { Visual representations of the \ha\ variability as a function of both spectral type and \vsini\ is} shown in Fig. \ref{fig:hafluxes}.

{
For the most stable star in the sample, GJ 1061, \ha\ is barely discernible, with variability of \hbox{$\sim$\,4 per cent} of the normalised continuum. Both GJ 1002 and GJ 3128 show \ha\ that is also filled in but with variability at the 20 per cent level. On the other hand, the M6.5V to M9 targets all show \ha\ in emission that varies considerably (see values in Table\ref{tab:activity}). The notable targets, however, are those exhibiting significant rotation, with \ha\ in strong emission, namely \hbox{LP 412-31}, \hbox{GJ 3146}, \hbox{LP 759-25} and \hbox{GJ 3076}. These targets possess the highest rotation in our sample, with \vsini\ values of 12, 12.4, 13 \& \hbox{17.1 \kms}\ respectively. GJ 3076 shows the least variability, indicative of saturation, while LP412-31 (with the highest measured EW) is also only moderately variable. { \cite{bell12halpha} also made this observation for the} complete M spectral range (M0V\,-\,M9V). They attributed this phenomenon to the higher level of persistent emission requiring significant heating (flaring) events to give a measurable change in emission.}

\cite{mohanty03activity}, \cite{west04halpha} and more recently \cite{reiners09volume1,reiners10activity} have studied rotation and activity across the M dwarf spectral class. By observing large samples, these studies indicated trends with chromospheric activity and \vsini. \cite{west04halpha} studied 8000 spectra of low mass stars from the Sloan Digital Sky Survey and found that \hbox{64\,-\,73 per cent} of M7V\,-M8V stars were active. { Here, although} our sample is small, we see considerable variability in { any specific object}. Hence, for the more active targets, a single snapshot observation is not necessarily representative of the mean activity level for that particular star. The trend first noted by \cite{mohanty03activity} and further quantified in \cite{reiners10activity}, suggests that \ha\ emission occurs at lower rotation rates in the later M stars. This is also apparent in our sample, where the M5V\,-M6V targets with slow rotation $\leq 5$ \kms, do not on the whole show a strong \ha\ line, whereas the M6.5V\,-\,M8V targets all possess significant \ha\ emission and variability for the similar rotation velocities. The sudden fall in L$_{\rm H_{\alpha}}$/\,L$_{\rm bol}$ noted by \cite{mohanty03activity} is seen in our latest targets, which despite similar rotation velocities of \hbox{6 \& 8 \kms}, show both the smallest EW$_{\rm {H_\alpha}}$ and log$_{10}$(L$_{\rm H_{\alpha}}$/\,L$_{\rm bol}$) { values}. Our findings are thus in keeping with the late spectral type activity frequency plots of \cite{reiners10activity} (see \hbox{their Fig. 7)}. 

\subsubsection{Morphology of \ha\ emission line}
\protect\label{section:halpha_morphology}

{ We make an additional observation regarding the shape of the \ha\ line, that may be applicable to stars (or subset populations of stars), such as the latest M dwarfs, where \ha\ is always seen in emission.} The exact morphology of the line appears to vary, with the emission profiles for some objects appearing to exhibit more pronounced double horned peaks than others. Further investigation of the detailed shape of \ha\ is warranted when it is realised that this shape is typical of emission from time varying circumstellar material at high stellar latitude. For example, \cite{barnes01aper} observed variability of \ha\ emission in the low axial inclination G8V \hbox{$\alpha$}\ Persei star AP 149, attributing it to a prominence system. A Doppler tomogram, derived using the code developed by \cite{marsh88tomo}, enabled four main emitting regions, located at and beyond co-rotation, to be inferred. While this technique requires sufficient velocity resolution to enable such a study, asymmetric variability of \ha\ emission may well be measurable in more slowly rotating stars. We find such variability at the 1\,-\,2 per cent level in the Proxima Centauri observations, with a trend suggesting a period that is greater than the five day time scale of the observations. With prolonged monitoring, the rotation period of stars that show \ha\ in strong emission may thus be estimated, while the exact shape of the emission (the prominence of the horns) may change with inclination angle.

\subsubsection{\ha\ and \vsini\ as a proxies for RV precision in late M stars}
\protect\label{section:halpha_proxy}

{ 
The upper panel of Fig. \ref {fig:activity_vs_rms} is a plot of \vsini\ vs r.m.s. (stellar line corrected BIS) values in this paper, illustrating the importance of \vsini\ in limiting the attainable precision as might intuitively be expected. We note that \hbox{2MASS J03341218-4953322}\ attains a precision that is greater than photon statistics predict (i.e. lower r.m.s.). This is probably a statistical effect that could potentially effect any small sample of observations. The contours plotted in Fig. \ref {fig:activity_vs_rms} were estimated by \cite{barnes13ropacs} using Monte-Carlo simulations with an M6V model atmosphere \citep{brott05}, while the increased number of opacities in an M9V star would lead us to expect a lower achievable precision. The Pearson correlation coefficient, $r$, gives an indication of the correlation. For \vsini\ vs r.m.s., we find \hbox{$r = 0.74$}, indicating a strong positive correlation.  The slope of the correlation itself is important when using \vsini\ as an indicator of expected precision. The discrepancy from the photon noise limited precision is greatest for the stars with the highest \vsini\ values, as we noted for the most rapid rotators in \S \ref{section:discussion}. Relying on \vsini\ to obtain an estimate of r.m.s. may therefore lead to an underestimation of the stellar jitter.

In Fig. \ref {fig:activity_vs_rms} (middle panel), the spectral type vs r.m.s. is plotted. Clearly the correlation with spectral type is weak, where we find $r = 0.04$. If we instead consider log$_{10}$(L$_{\rm H_{\alpha}}$/\,L$_{\rm bol}$) as an indicator of r.m.s., as plotted in Fig. \ref {fig:activity_vs_rms} (bottom panel), we again see a clear trend. The Pearson correlation coefficient for the lower and upper log$_{10}$(L$_{\rm H_{\alpha}}$/\,L$_{\rm bol}$) values are $r = 0.76$ \& $0.82$ respectively. Considering upper and lower limits together, we obtain $r = 0.77$. The significance of the trend of \vsini\ vs r.m.s. compared with log$_{10}$(L$_{\rm H_{\alpha}}$/\,L$_{\rm bol}$) vs r.m.s. across our sample are thus comparable. It would appear that the absorption lines of late-type stars are significantly affected by magnetic activity, especially when moderate rotation of \vsini\,$\sim$\,10 \kms\ and above is observed. Although relying on \vsini\ to estimate r.m.s. may underestimate the jitter in this regime, the use of \ha\ emission level instead removes the rotation dependence. 
}

\subsubsection{Ca {\sc ii} 8662 \AA\ activity and correlation with \ha\ variability}
\protect\label{section:calcium_activity}

{ The Ca {\sc ii} H \& K lines have regularly been monitored in F-M type stars for many years (e.g. \citealt{wilson78,baliunas95}) since their emission cores show strong variability connected with stellar magnetic activity. The $S$ index measured from the H \& K lines \citep{baliunas95} is known to be a general indicator of activity as it is related to the area and the strength of magnetic activity on a star \citep{schrijver89}. Stars with low log\,$R'_{\rm HK}$ indices (the fraction of a star's luminosity in the \hbox{Ca {\sc ii} H \& K lines}) are generally selected for precision radial velocity searches for planets (e.g. \citealt{wright04activity}). The role of Ca {\sc ii} H \& K excess emission and it's relationship with jitter in the large sample of the California Planet Search has been studied by \cite{isaacson10} for instance. In the subset of their sample that includes the latest stars (early M dwarfs), a noise floor is seen with evidence for a trend that increases with activity, as discussed in \S \ref{section:halpha_proxy} above.

Although the Ca {\sc ii} H \& K lines are very strong and easily accessible for F-K type stars observed with most high resolution spectrometers, the flux at blue wavelengths, especially by mid M spectral type is too low to enable sufficient S/N to be attained during typical observations. Other Ca {\sc ii} lines that are sensitive to chromospheric activity, such as the so called infrared \hbox{Ca {\sc ii}} triplet, are however observed in the wavelength regime in which our survey operates. Of the infrared \hbox{Ca {\sc ii}} triplet lines at 8498, 8542 \& 8662 \AA, the latter line appears the least blended. Hence we chose to illustrate the non-LTE behaviour (i.e. potential emission in the core) of this line in Fig. \ref{fig:ca_ha}. The line becomes indistinct, through blending with other lines, in the later spectral types in our sample. In our ROPS sample, variability above the noise level can be discerned in Fig. \ref{fig:ca_ha}, notably for \hbox{GJ 3076} and \hbox{LP 412-31}. The clearest variation in this line is seen with LP 759-25 (similar variability is also seen, but not plotted, in the other two infrared Ca {\sc ii} lines). 

It appears that variability in the  Ca {\sc ii} 8662 \AA\ line is only easily discerned for very strong flares. \cite{fuhrmeister07cnleo} observed the behaviour of the \hbox{Ca {\sc ii}} triplet lines for the flaring M5.5 dwarf \hbox{CN Leo}, noting the correlation with other chromospheric lines. In the case of Proxima Centauri, \cite{fuhrmeister11proxcen} presented {\sc uves} observations of \hbox{Ca {\sc ii} H\&K}, \ha\ alongside optical lightcurves (obtained with the blue exposure meter of {\sc uves}). The \hbox{Ca {\sc ii}} triplet lines were not discussed in their study, however simultaneous observations with XMM-Newton, covering the \hbox{0.2\,-\,10 keV} range and the U band \hbox{(300\,-\,390 nm)} were presented. We have also included \ha\ and Ca {\sc ii} 8662 \AA\ in Fig. \ref{fig:ca_ha} to demonstrate the range of variability seen over all observations of the 2009 data. Profiles are included for the minimum, mean and maximum states, with the latter corresponding to the strongest flaring event on the final night. Although the correlation between \ha\ variability and the infrared Ca {\sc ii} triplet variability was not included in the study by \cite{fuhrmeister11proxcen}, their Figs. 1\,-\,3 showed a strong correlation between \ha\ and Ca {\sc ii} H\&K (albeit at a lower observation cadence necessitated by the longer exposure times required in the blue arm of {\sc uves} with and M6V star. Since the infrared Ca {\sc ii} triplet lines are heavily blended, we have determined the variability in Ca {\sc ii} 8662 \AA\ by subtracting the mean spectrum (derived from all observations). The relative EW was then measured for each observation. We find the correlation between \ha\ and Ca {\sc ii} 8662 \AA\ EW values is very strong, with Pearson correlation coefficients of $r = 0.92, 0.94$ \& $0.91$ for each of the 3 nights. The Ca {\sc ii} triplet lines are thus potentially useful for identifying strong flaring events, although the strength of \ha\ makes it the more useful line for activity monitoring in the first instance.  
}


\subsubsection{Selection of RVs based on activity events}
\protect\label{section:activity_events}

{ A very large flare was observed during the observations of CN Leo \citep{reiners09flare} that lead to 660 \ms\ deviation from the other radial velocities that were measured with $\sim 10$ \ms\ precision. For all other flare events resulting in log$_{10}$(L$_{\rm H_{\alpha}}$/\,L$_{\rm bol}$) changes of $\leq$\,0.4 dex, \cite{reiners09flare} found no RV variability at the 10 \ms\ precision of the observations. The conclusion from that study is that only the very strongest flares, that are easily identifiable in spectra, affect RVs at the level of \hbox{$\sim$500 \ms}. The large flare on Proxima Centauri on 2009 March 14 resulted in a change of \ha\ emission of \hbox{0.33 dex} (comparing the immediate pre-flare and maximum flare EWs). Our r.m.s. precision on \hbox{2009 March 14} was \hbox{5.84 \ms} (full 4-parameter correction). However excursions of up to \hbox{20 \ms}\ can occasionally be seen in the bottom panels of Fig. \ref{fig:proxcen_rvs} that do not necessarily coincide with the flaring events presented in \cite{fuhrmeister11proxcen}. Before and after the sudden rise in \ha\ emission corresponding to the large flare on 2009 March 14, the RVs appear to be relatively stable (Fig. \ref{fig:proxcen_rvs}, bottom right panel) in the the \hbox{t $\sim$ 4.23\,-\,4.25 day} and \hbox{4.28\,-\,4.30 day} regions. However from \hbox{t $\sim$ 4.26\,-\,4.28}, there is systematic RV deviation of \hbox{$\sim$ 20 \ms}\ coinciding with the onset and peak of the flare. It is difficult to determine whether the flare is responsible since systematic deviations of similar magnitude occur on \hbox{2009 March 12} and earlier on \hbox{2009 March 14}. No strong flare counterpart is seen in the other activity indicators presented in \cite{fuhrmeister11proxcen} for these RV deviations. The correlation coefficients between \ha\ EWs and RVs for the complete Proxima Centauri data set is $r = -0.09$. For the region where \ha\ EW increases and the RVs show the tentative peak (t = 4.255\,-\,4.276 day), $r = -0.28$, indicating a weak negative correlation. The behaviour of \ha\ is more complex however during this strong flare cascade. \hbox{Fig. 3} in \cite{fuhrmeister11proxcen} shows that the  H$\epsilon$ and \hbox{H I 3770 \AA}\ lines exhibit a more clearly defined peak (i.e. a sharper decline after the sharp rise at XMM-Newton wavelengths) that may indicate a higher correlation with the RVs. In conclusion, it is not clear that the strong flare really impacted on our RVs in this case and there is no evidence that any of the other flare events affected the RVs on Proxima Centauri at the 4\,-\,6 \ms\ level during the three nights of the observations.
}

{ 
Although the evidence suggests that moderate flaring does not affect RV measurements at $\geq 10$ \ms\ on {\em slow} rotators, it is not clear whether this is also true for {\em moderate} rotators. In this instance, activity related transients might be more clearly resolved owing to Doppler broadening of the lines. If we remove the observation of LP 759-25 \hbox{(\vsini = 13.7 \kms)}, which shows \ha\ in strong emission and evidence for a large flare (in both \ha\ and Ca {\sc ii} 8662 \AA), the measured r.m.s. reduces from 79.9 \kms\ (line bisector corrected) to 29.7 \kms. While this represents a dramatic improvement, and is now twice our predicted photon noise limited precision of \hbox{15 \ms}\ (see earlier discussion in this section), it is difficult to determine the significance given that the r.m.s. values are based on only 4 and 3 observations alone respectively.}


It is thus clear that more observations are needed { for each star because if late-M stars are moderate rotators on average that show modulated activity}, then measuring precise radial velocities at the sub-10 \ms\ level will prove extremely challenging. { De-trending of RVs using activity indicators generally utilises of order 20\,-\,30 epochs, at which stage planetary signals can be well characterised (e.g. see \citealt{bonfils13mdwarfs}). Monitoring of activity indicators for strong flaring events in late M dwarfs is also essential. For CN Leo \citep{reiners09flare} only $\sim$\,$4$ per cent of observations were affected by a strong flare. We also see very tentative evidence (with weak correlation), for RVs affected by flare activity (at the 20 \ms\ level) in $\sim$\,$4$ per cent of the Proxima Centauri observations.}





\begin{figure}
\begin{center}
\includegraphics[height=85mm,angle=270]{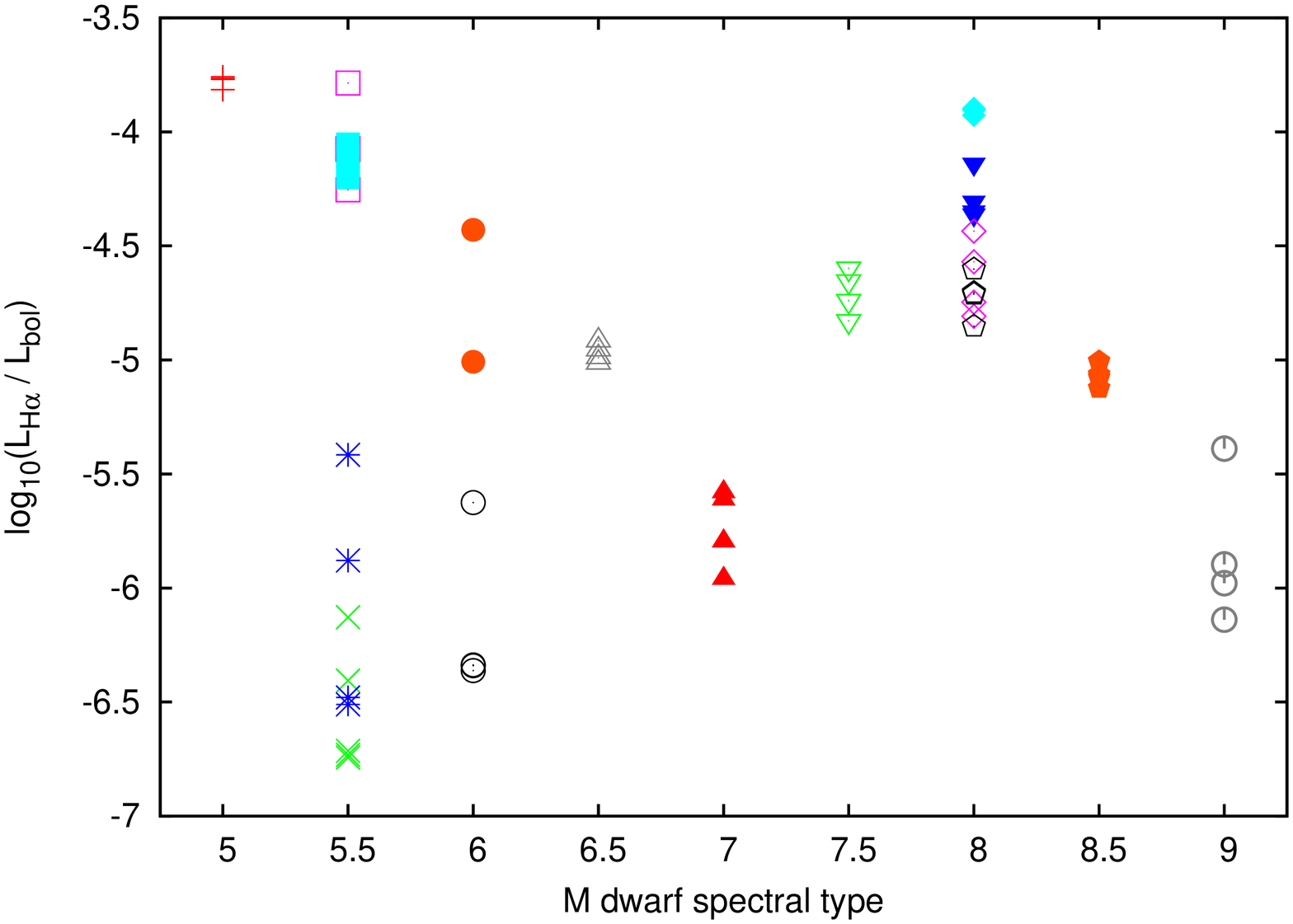} \\
\includegraphics[height=85mm,angle=270]{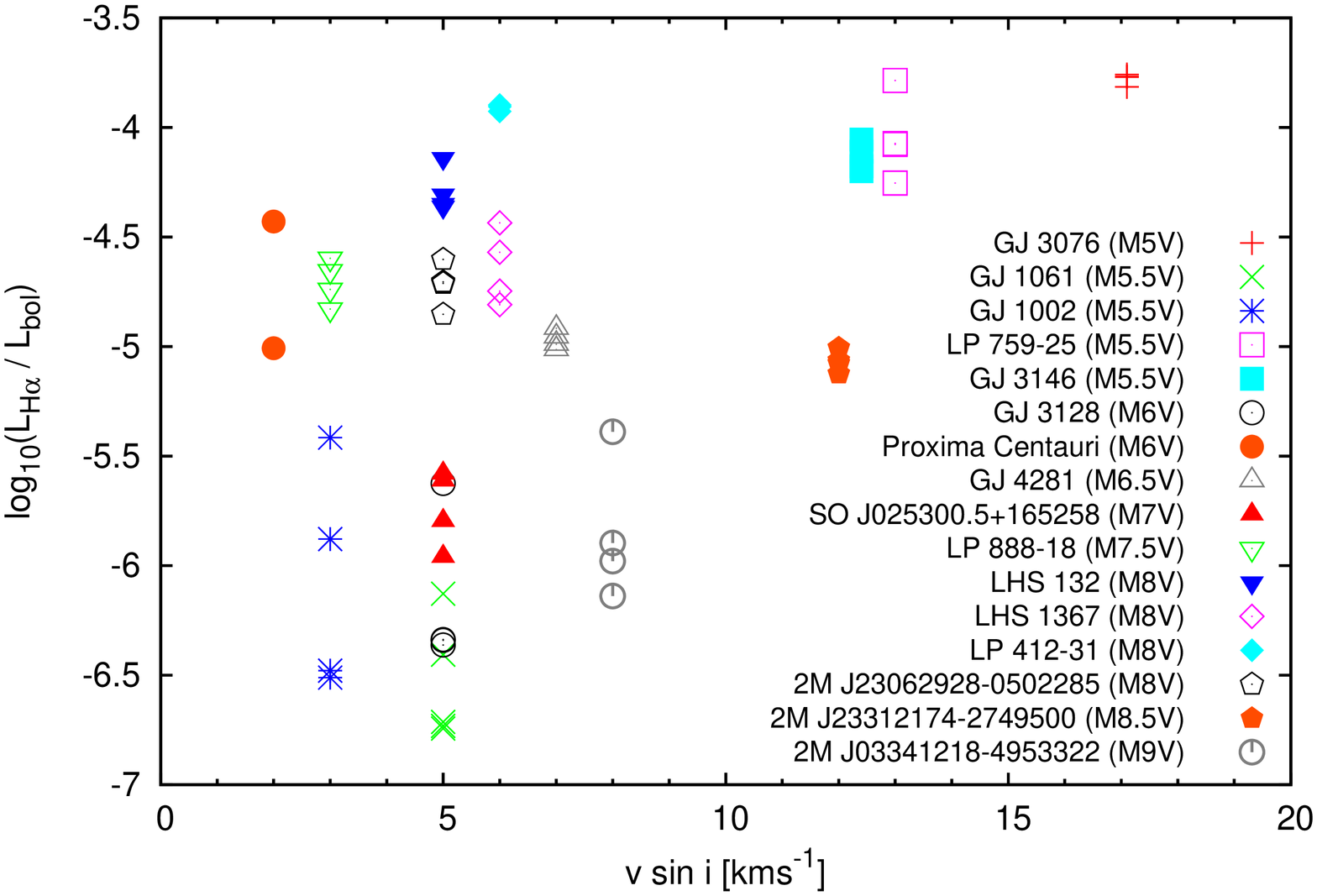} \\
\end{center}   
\caption{Activity, log$_{10}$(L$_{\rm H_{\alpha}}$/\,L$_{\rm bol}$) as a function of spectral type (top) and \vsini\ (bottom) for the 15 ROPS targets and Proxima Centauri. The symbols and colours for each object are indicated in the lower panel key and apply to both plots. The earliest star with significant rotation exhibits the highest log$_{10}$(L$_{\rm H_{\alpha}}$/\,L$_{\rm bol}$), while down to M8V, significant activity variation is seen at more moderate rotation speeds. Except for Proxima Centauri, the slowly rotating M5.5V\,-M6V stars show little \ha\ activity, while the latest stars in the sample (M8.5V \& M9V) are also less active.}
\protect\label{fig:hafluxes}
\end{figure}

\begin{figure}
\begin{center}
\includegraphics[height=83mm,angle=270]{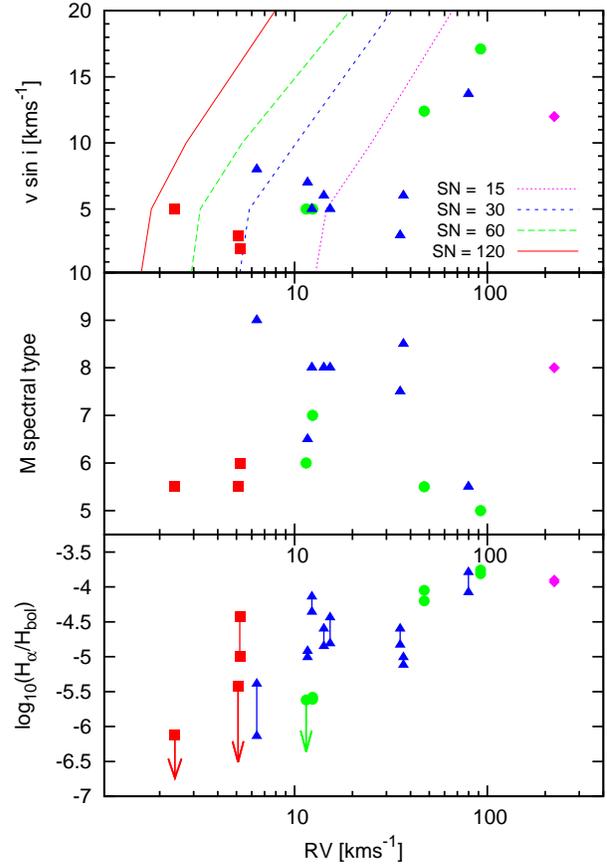} \\
\end{center}   
\caption{Key stellar parameters plotted against r.m.s. (line BIS corrected) for the ROPS targets and Proxima Centauri. The plots are of \vsini\ vs r.m.s. (top), spectral type vs r.m.s. (middle) and activity (log$_{10}$(L$_{\rm H_{\alpha}}$/\,L$_{\rm bol}$)) vs r.m.s. (bottom). The symbols and colours used in all panels denote the S/N ratios or S/N ratio intervals for each observed target: \hbox{$0$ $\leq$ S/N $<$ 15} (red squares), \hbox{$15$ $\leq$ S/N $<$ 30} (green circles), \hbox{$30$ $\leq$ S/N $<$ 60} (blue triangles), \hbox{S/N $\geq$ $100$} (magenta diamonds). Similarly, photon noise limited contours from Barnes (2013) are plotted in the top panel for S/N = 15, 30, 60 and 120 respectively (red/solid, green/long-dash, blue/short-dash, magenta/dotted). The stars with the highest \vsini\ values are most discrepant from the photon noise limited case, indicating the importance of activity as an indicator of expected precision. Maximum and minimum values of \ha\ luminosity, as given in Table 2, are plotted as circles connected by a line for each star in the bottom panel. For GJ 1061, 1002 \& 3128, very small line equivalent widths were found for some or all (GJ 1061) phases. The arrow head indicates that the lowest \ha\ luminosity is a sensitivity limit, and equal to the equivalent width uncertainty. }
\protect\label{fig:activity_vs_rms}
\end{figure}

\section{Conclusion and future prospects}
\protect\label{section:conclusion}

With careful wavelength calibration we have demonstrated that \hbox{{ 2.4} \ms}\ precision can be achieved with {\sc uves} operating in the red part of the optical. Since we have so far only obtained 3\,-\,5 radial velocity measurements per star spanning 6 nights, further observations are required before any potential planetary signals can be discerned. However, under the assumption that the current r.m.s. estimates are representative of a larger set of observations, and by considering our most stable targets that exhibit r.m.s. $<$\,$16$ \ms, we can rule out the presence of planets with \hbox{$m_p$\,sin\,$i$ $>$\,10 \mearth}\ in \hbox{0.03 AU} orbits. Extending this to include the less stable stars, our observations do not support evidence for planets { more} massive than \hbox{$m_p$\,sin\,$i$ = 0.5 M$_{\rm J}$} at \hbox{0.03 AU}. Fig. \ref{fig:rops_paramspace_frompng} illustrates the late M parameter space of this investigation, presenting our ROPS targets, the early M dwarfs, and all planets detected with radial velocities, orbiting stars up to 2 \msun. The RVs corrected with the stellar line BIS, are plotted as upper limits, and at present are only intended to illustrate the sensitivities achieved with our survey. 

Even a modest survey, targeting relatively small numbers of \hbox{M dwarfs}, leads us to expect significant numbers of low-mass planets, following the findings of recent studies (e.g. \cite{bonfils13mdwarfs,kopparapu13,dressing13}). It is vitally important however that if precisions of a few \ms\ are to be obtained in a large sample of stars, the activity must be clearly characterised and understood. This study was motivated by the uncharted late M dwarfs, hence our inclusion of target stars that were typical, in terms of activity and rotation. Even with the few observations presented here it is clear that any future surveys must carefully select targets that do not bias the sample. In other words, selection of only slowly rotating late M dwarfs might lead to observation of { predominantly low axial inclination systems ($i$\,$\ll$\,$90$\degs) that} are not as favourable for RV planet detection. { For instance, if the mean axial inclination of a stellar population is assumed to be 45\degs, and if the mean \vsini\ is found to be \hbox{8 \kms} for a typical M6V \citep{jenkins09mdwarfs}, selection of stars with \hbox{\vsini\ $< 5$ \kms}\ will lead to a sample biased to a mean axial inclination of \hbox{$i$\,$<$\,$26$\degs}. The problem becomes worse for an M9V star with a typical \hbox{\vsini\ = 15 \kms}. Here we expect observation of stars with \hbox{\vsini\ $< 5$ \kms}\ will lead to a sample with a mean axial inclination of \hbox{$i$\,$<$\,$14$\degs}\ (i.e. very close to pole-on).} Conversely, we find that significant \vsini\ leads to higher r.m.s., and most importantly that the these r.m.s. values are significantly above the photon noise limited precision at the observed \vsini\ and S/N ratios. \ha\ luminosity, log$_{10}$(L$_{\rm H_{\alpha}}$/\,L$_{\rm bol}$), shows a clear trend with r.m.s. as discussed in \S \ref{section:halpha_proxy} and illustrated in \hbox{Fig. \ref {fig:activity_vs_rms}}. The implication that late type stars are significantly spotted, and hence exhibit time varying line distortions, suggests that ways of mitigating the effects of the resulting ``jitter'' are important for this class of stars.

In this paper, we used a standard and straightforward bisector span analysis to de-trend the data. Only further observations will enable this procedure to be fully validated. At the same time, simple BIS analysis is not able to properly remove any magnetically induced RV signatures to the photon-noise level, especially for stars with moderate rotation. We will investigate this in a future publication, but { our preliminary simulations indicate} that BIS analysis is optimal for a narrow range of rotation velocities. Any starspot distributions are also an important consideration. In \cite{barnes11jitter} for example, we assumed the effects of randomly distributed spots, but did not try to remove their influence on the radial velocity jitter. It is not at all clear that random spots, resulting from a fully convective turbulent dynamo, are the predominant spot pattern on active late M spectral types. By mid-M when stars become fully convective there is evidence that magnetic fields become dipolar \citep{donati08mdwarfs,morin08mdwarfs}. Doppler images are one means of characterising spot patterns, but photospheric brightness images have currently only been derived for early-M stars \citep{barnes01mdwarfs,barnes04hkaqr,phanbao09mdwarfs}.

Observing strategies are also an important consideration when trying to mitigate any starspot effects. \cite{moulds13} has found that starspot jitter can largely be removed by modelling starspot effects on the line profile. Hence intensive spectroscopic observations of late M targets may be necessary to enable more effective removal of activity signatures. Fortuitously, M6V\-,\,M9V planets are expected { in close orbit about} their parent stars \citep{bonfils13mdwarfs}, which as already noted, lead us to expect \hbox{6\,-\,11 day} orbits at the centre of the continuous habitable zone \citep{kopparapu13hz}. Hence observations over week-month long timescales over which starspot groups are stable \citep{goulding12wfcam}, should enable good sampling of M dwarf planet orbits while simultaneously providing the observations that could help remove activity jitter. In addition, the use of Bayesian techniques to search for low amplitude signals in noise { enables recovery of} radial velocity signatures in only a few epochs. Indeed we find low amplitude signals in the {\sc harps}\ early-mid M dwarf sample \citep{tuomi13mdwarfs} that indicate RV signals are abundant, with occurrence rates of $0.06^{+0.11}_{-0.03}$ for \hbox{$3$ \mearth\ $\leq$\,$m_p$\,sin\,$i$ $\leq$ $10$ \mearth} in $1$\,-\,$10$ day orbits, increasing to  $1.02^{+1.48}_{-0.69}$ for 10\,-\,100 day planets (i.e. an upper limit of greater than one planet per star). Moreover, the estimated habitable zone occurrence rate for \hbox{$3$ \mearth\ $\leq$\,$m_p$\,sin\,$i$ $\leq$ $10$ \mearth}, is found to be \hbox{$\eta_\oplus$ = 0.16\,-\,0.24}. By extrapolation from early M dwarf observations, we expect late M dwarf planet frequencies to peak in shorter orbits, continuing the trend of semi-major axis distribution vs stellar mass noted by \cite{currie09}. For example, the \hbox{33 day}, \hbox{0.135 AU} orbit \citep{kopparapu13hz} of a HZ planet hosted by \hbox{0.3 \msun}\ early M dwarf would reduce to an 11 day, 0.045 AU orbit for the sample planet hosted by a \hbox{0.1 \msun}\ star. Further to the above argument, this illustrates that short observing campaigns should quickly uncover significant signals for surveys that enable few \ms\ precision to be attained. The search for low-mass planets orbiting the lowest mass stars is thus a challenging but achievable goal with current estimates leading us to expect a host of interesting planets in the near future.
                                                                        
\section*{Acknowledgments}
{ We would like to thank the anonymous referee for the suggested amendments and for careful reading of the manuscript}. JB gratefully acknowledges funding through a University of Hertfordshire Research Fellowship. JSJ acknowledges funding by Fondecyt through grant 3110004 and partial support from Centro de Astrof\'{i}sica FONDAP 15010003, the GEMINI-CONICYT FUND and from the Comit\'{e} Mixto ESO-GOBIERNO DE CHILE. SVJ acknowledges research funding by the Deutsche Forschungsgemeinschaft (DFG) under grant SFB 963/1, project A16. DM and PA gratefully acknowledge support by the FONDAP Center for Astrophysics 15010003, the BASAL CATA Center for Astrophysics and Associated Technologies PFB-06, and the MILENIO Milky Way Millennium Nucleus from the Ministry of Economy’s ICM grant P07-021-F. AJ acknowledges support from FONDECYT project 1130857, BASAL CATA PFB-06, and the Millennium Science Initiative, Chilean Ministry of Economy (Millennium Institute of Astrophysics MAS and Nucleus P10-022-F). PR also acknowledges FONDECYT project 1120299. During the course of this work, DJP and MT were supported by RoPACS, a Marie Curie Initial Training Network funded by the European Commission’s Seventh Framework Programme. JB, JSJ, DJP and SVJ have also received travel support from RoPACS during this research. This paper includes data gathered with the 6.5 meter Magellan Telescopes located at Las Campanas Observatory, Chile.

\begin{figure}
\begin{center}
\includegraphics[height=60mm,angle=0,]{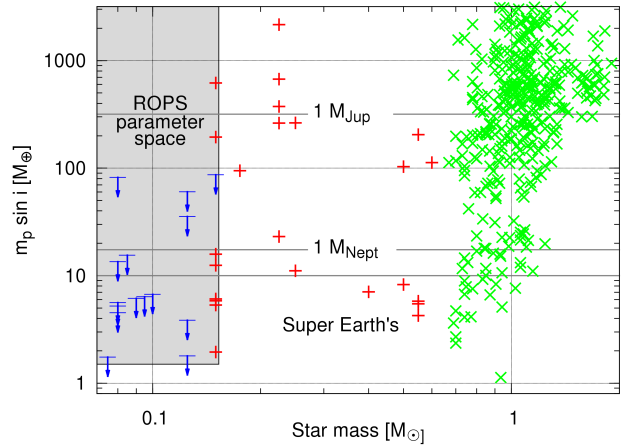} \\
\end{center}   

\caption{Unique parameter space explored by ROPS and current r.m.s. limits (blue filled circles with downward arrows indicating upper limits). Also shown are the M dwarfs with known planetary companions (red crosses) and all known planets with radial velocity measurements and hence $m_p$\,sin\,$i$ determinations.}
\protect\label{fig:rops_paramspace_frompng}
\end{figure} 


\newpage
\appendix
\section{Radial velocities}
\protect\label{section:appendix_rvs}

The radial velocity measurements are calculated via the procedures described in this paper, using line lists derived from \hbox{GJ 1061 (M5.5V)} and \hbox{LHS 132 (M8V)}. Two observations were made of GJ 1061 at high resolution (0.4\arcsec\ slit) and were co-aligned, co-added and normalised to a value of 1.0 to obtain a template for empirical determination of the line list. Similarly all four observations of LHS 132 were aligned to the first observation (0.8\arcsec\ slit) and the resulting template normalised to a value of 1.0. The line lists were derived by identifying the minima of absorption features in the templates and fitting quadratics to the three lowest values in each absorption line. The line depth and wavelength of each line were thus recorded.

Since we did not observe a radial velocity standard, all radial velocities are determined relative to the template used for deconvolution. The mean heliocentrically corrected velocity of the radial velocity observations of each of the template stars is first determined. For GJ 1061, we find \hbox{$\Gamma_{\rm GJ1061} = 14499.07$ \ms}\ and for \hbox{LHS 132}, \hbox{$\Gamma_{\rm LHS132} = 18661.17$ \ms}. In other words, the RVs listed for \hbox{GJ 1061} (Tables \ref{tab:appendix_table_proxcen} \& \ref{tab:appendix_table_gj1061}) and \hbox{LHS 132} (Table \ref{tab:appendix_table_lhs132}), have been determined by subtracting the indicated \hbox{$\Gamma_{\rm GJ1061}$} and \hbox{$\Gamma_{\rm LHS132}$} values. The velocities relative to the reference frame of \hbox{GJ 1061} can thus be obtained from columns 2, 4, 5 \& 6 of Table A1 and columns 2, 4 \& 5 of Table A2. Similarly the velocities relative to the reference frame of \hbox{LHS 132} can be obtained from columns 2, 4 \& 5 of Table A3.

For all other targets, we indicate the deconvolution template used (either GJ 1061 or LHS 132) and the value $\Gamma_{*}$ (where * denotes the star) that must be added to the tabulated velocities in order to place them in the reference frame of that template. We tabulate the $\Gamma_{*}$ subtracted values (i.e. zero mean) for consistency with Fig. \ref{fig:rvs} which makes the RV variability easier to discern.


\setcounter{table}{0}
\begin{table*}
\caption{Observation times and velocities for Proxima Centauri. The $\Gamma_{\rm ProxCen}$ velocity must be added to the individual velocities to transform them into the reference frame of GJ 1061 (from which the deconvolution template was derived). Columns 1\,-\,3 give the Julian date, the velocities before atmospheric correction relative to $\Gamma_{\rm ProxCen}$ (as presented in the upper panels of Fig. \ref{fig:proxcen_rvs}) and the propagated uncertainty for each observation. The velocities after the atmospheric correction is applied to each night individually are given in columns 4 (I corr) and for all nights together in column 5 (A corr). See \S 4.3 for details. }
\begin{tabular}{ccccc}
\hline
JD              &  RV	     & RV Error          &        RV     &      RV       \\
                 &[ms$^{-1}]$& [ms$^{-1}]$       & [ms$^{-1}]$   & [ms$^{-1}]$   \\
                 &  No corr  &                   & I corr        & A corr        \\
\hline
\multicolumn{5}{c}{Proxima Centauri ($\Gamma_{\rm ProxCen} = -22344.98 $ \ms) } \\
\hline
 2454900.134561   &      44.24   &       7.08   &       0.00   &       0.00   \\ 
 2454900.136059   &      36.43   &       6.02   &       3.69   &      -5.47   \\ 
 2454900.137557   &      41.09   &       6.07   &      -3.85   &     -12.82   \\ 
 2454900.139055   &      36.46   &       6.84   &       1.05   &      -7.72   \\ 
 2454900.140552   &      43.11   &       6.56   &      -3.33   &     -11.90   \\ 
 2454900.142048   &      35.45   &       6.73   &       3.58   &      -4.80   \\ 
 2454900.143546   &      34.04   &       5.99   &      -3.85   &     -12.02   \\ 
 2454900.145049   &      41.68   &       7.29   &      -5.03   &     -13.00   \\ 
 2454900.146545   &      35.18   &       5.79   &       2.84   &      -4.93   \\ 
 2454900.148045   &      36.54   &       6.37   &      -3.44   &     -11.01   \\ 
 2454900.149547   &      43.95   &       6.55   &      -1.86   &      -9.23   \\ 
 2454900.151045   &      49.28   &       7.01   &       5.77   &      -1.40   \\ 
 2454900.152543   &      33.51   &       6.28   &      11.30   &       4.33   \\ 
 2454900.154043   &      36.71   &       6.40   &      -4.26   &     -11.03   \\ 
 2454900.155546   &      35.31   &       6.49   &      -0.86   &      -7.42   \\ 
 2454900.157045   &      38.47   &       5.80   &      -2.09   &      -8.44   \\ 
 2454900.158541   &      37.84   &       5.89   &       1.26   &      -4.89   \\ 
 2454900.160038   &      48.72   &       6.73   &       0.83   &      -5.12   \\ 
 2454900.161536   &      38.19   &       6.22   &      11.87   &       6.13   \\ 
 2454900.163032   &      41.80   &       6.52   &       1.53   &      -4.02   \\ 
 2454900.164534   &      35.23   &       6.28   &       5.31   &      -0.03   \\ 
 2454900.166037   &      42.76   &       6.41   &      -1.10   &      -6.24   \\ 
 2454900.167537   &      40.66   &       6.51   &       6.60   &       1.66   \\ 
 2454900.169039   &      28.53   &       6.35   &       4.66   &      -0.07   \\ 
 2454900.170536   &      37.53   &       7.09   &      -7.31   &     -11.85   \\ 
 2454900.172035   &      31.18   &       6.29   &       1.85   &      -2.49   \\ 
 2454900.173536   &      35.62   &       7.02   &      -4.37   &      -8.50   \\ 
 2454900.175038   &      36.42   &       6.43   &       0.24   &      -3.70   \\ 
 2454900.176538   &      29.02   &       5.84   &       1.17   &      -2.56   \\ 
 2454900.178037   &      35.78   &       6.73   &      -6.09   &      -9.62   \\ 
 2454900.179535   &      29.29   &       5.49   &       0.81   &      -2.54   \\ 
 2454900.181036   &      32.33   &       5.09   &      -5.56   &      -8.70   \\ 
 2454900.182533   &      33.84   &       5.82   &      -2.38   &      -5.33   \\ 
 2454900.184029   &      29.40   &       6.15   &      -0.74   &      -3.50   \\ 
 2454900.185530   &      33.65   &       6.10   &      -5.04   &      -7.62   \\ 
 2454900.187028   &      34.94   &       6.04   &      -0.67   &      -3.05   \\ 
 2454900.188531   &      29.71   &       6.75   &       0.74   &      -1.45   \\ 
 2454900.190033   &      29.76   &       5.55   &      -4.36   &      -6.37   \\ 
 2454900.191532   &      28.58   &       5.23   &      -4.19   &      -6.02   \\ 
 2454900.193034   &      29.90   &       5.23   &      -5.26   &      -6.90   \\ 
 2454900.194535   &      34.38   &       6.42   &      -3.82   &      -5.27   \\ 
 2454900.196035   &      27.25   &       4.95   &       0.77   &      -0.51   \\ 
 2454900.197537   &      36.39   &       4.82   &      -6.25   &      -7.35   \\ 
 2454900.199039   &      33.26   &       3.40   &       2.99   &       2.07   \\ 
 2454900.200540   &      35.44   &       3.70   &      -0.02   &      -0.76   \\ 
 2454900.202039   &      26.18   &       4.51   &       2.25   &       1.69   \\ 
 2454900.203541   &      32.65   &       5.06   &      -6.89   &      -7.29   \\ 
 2454900.205041   &      32.21   &       5.50   &      -0.31   &      -0.54   \\ 
 2454900.206542   &      35.40   &       4.71   &      -0.66   &      -0.72   \\ 
 2454900.208043   &      30.21   &       5.82   &       2.64   &       2.75   \\ 
 2454900.209541   &      35.96   &       5.89   &      -2.44   &      -2.18   \\ 
 2454900.211041   &      35.94   &       5.21   &       3.39   &       3.82   \\ 
 2454900.212542   &      39.80   &       5.89   &       3.47   &       4.06   \\ 
 2454900.214043   &      34.54   &       6.54   &       7.42   &       8.17   \\ 
 2454900.215541   &      35.20   &       5.50   &       2.26   &       3.15   \\ 
 2454900.217036   &      36.69   &       5.30   &       3.02   &       4.06   \\ 
\hline
\end{tabular}
\protect\label{tab:appendix_table_proxcen}
\end{table*}

\setcounter{table}{0}
\begin{table*}
\caption{Continued.}
\begin{tabular}{ccccc}
\hline
JD              &  RV	     & RV Error          &        RV     &      RV       \\
                 &[ms$^{-1}]$& [ms$^{-1}]$       & [ms$^{-1}]$   & [ms$^{-1}]$   \\
                 &  No corr  &                   & I corr        & A corr        \\
\hline
\multicolumn{5}{c}{Proxima Centauri ($\Gamma_{\rm ProxCen} = -22344.98 $ \ms) } \\
\hline
 2454900.218535   &      27.12   &       5.52   &       4.59   &       5.79   \\ 
 2454900.220034   &      31.95   &       5.11   &      -4.88   &      -3.54   \\ 
 2454900.221531   &      32.63   &       5.32   &       0.05   &       1.53   \\ 
 2454900.223029   &      35.98   &       5.25   &       0.82   &       2.44   \\ 
 2454900.224525   &      29.80   &       4.69   &       4.25   &       6.02   \\ 
 2454900.226026   &      38.48   &       5.54   &      -1.83   &       0.06   \\ 
 2454900.227527   &      31.88   &       6.18   &       6.93   &       8.97   \\ 
 2454900.229027   &      27.97   &       5.47   &       0.42   &       2.58   \\ 
 2454900.230527   &      31.13   &       5.96   &      -3.40   &      -1.10   \\ 
 2454900.232025   &      36.79   &       6.00   &      -0.15   &       2.27   \\ 
 2454900.233526   &      30.10   &       4.91   &       5.58   &       8.13   \\ 
 2454900.235026   &      34.16   &       4.39   &      -1.01   &       1.65   \\ 
 2454900.236526   &      29.70   &       4.18   &       3.13   &       5.92   \\ 
 2454900.238028   &      34.31   &       4.59   &      -1.25   &       1.65   \\ 
 2454900.239526   &      23.29   &       5.30   &       3.44   &       6.46   \\ 
 2454900.241028   &      23.64   &       4.22   &      -7.50   &      -4.37   \\ 
 2454900.242528   &      31.06   &       5.11   &      -7.06   &      -3.83   \\ 
 2454900.244029   &      35.21   &       5.88   &       0.45   &       3.79   \\ 
 2454900.245530   &      25.08   &       4.65   &       4.68   &       8.12   \\ 
 2454900.247028   &      29.84   &       5.80   &      -5.38   &      -1.84   \\ 
 2454900.248529   &      35.46   &       5.68   &      -0.52   &       3.10   \\ 
 2454900.250030   &      27.16   &       5.61   &       5.18   &       8.90   \\ 
 2454900.251535   &      34.25   &       5.72   &      -3.02   &       0.78   \\ 
 2454900.253037   &      30.44   &       5.28   &       4.13   &       8.02   \\ 
 2454900.254538   &      30.92   &       4.78   &       0.42   &       4.39   \\ 
 2454900.256037   &      29.30   &       5.70   &       0.97   &       5.03   \\ 
 2454900.257537   &      31.10   &       4.28   &      -0.56   &       3.57   \\ 
 2454900.259035   &      29.91   &       4.14   &       1.32   &       5.52   \\ 
 2454900.260533   &      29.89   &       4.59   &       0.21   &       4.48   \\ 
 2454900.262029   &      34.57   &       5.08   &       0.26   &       4.61   \\ 
 2454900.263528   &      32.81   &       5.56   &       5.04   &       9.44   \\ 
 2454900.265030   &      29.58   &       5.65   &       3.37   &       7.83   \\ 
 2454900.266528   &      31.84   &       5.79   &       0.22   &       4.74   \\ 
 2454900.268026   &      34.04   &       5.58   &       2.55   &       7.13   \\ 
 2454900.269527   &      27.31   &       6.03   &       4.85   &       9.47   \\ 
 2454900.271025   &      29.23   &       4.87   &      -1.81   &       2.86   \\ 
 2454900.272526   &      32.27   &       5.75   &       0.21   &       4.92   \\ 
 2454900.274028   &      33.35   &       5.18   &       3.32   &       8.08   \\ 
 2454900.275530   &      32.45   &       5.15   &       4.50   &       9.29   \\ 
 2454900.277030   &      31.66   &       5.46   &       3.67   &       8.50   \\ 
 2454900.278531   &      24.55   &       4.42   &       2.99   &       7.83   \\ 
 2454900.280027   &      17.96   &       4.24   &      -4.05   &       0.83   \\ 
 2454900.281529   &      27.33   &       4.25   &     -10.55   &      -5.65   \\ 
 2454900.283028   &      24.33   &       4.87   &      -1.08   &       3.83   \\ 
 2454900.284525   &      27.81   &       4.75   &      -4.01   &       0.93   \\ 
 2454900.286027   &      24.88   &       4.88   &      -0.43   &       4.52   \\ 
 2454900.287525   &      32.07   &       5.02   &      -3.27   &       1.68   \\ 
 2454900.289026   &      28.74   &       4.92   &       4.00   &       8.96   \\ 
 2454900.290527   &      31.17   &       4.81   &       0.76   &       5.72   \\ 
 2454900.292028   &      27.17   &       5.14   &       3.28   &       8.25   \\ 
 2454900.293530   &      26.43   &       5.00   &      -0.63   &       4.32   \\ 
 2454900.295028   &      26.09   &       5.17   &      -1.27   &       3.68   \\ 
 2454900.296528   &      21.98   &       3.80   &      -1.52   &       3.41   \\ 
 2454900.298028   &      19.80   &       4.03   &      -5.54   &      -0.62   \\ 
 2454900.299529   &      25.45   &       4.76   &      -7.63   &      -2.73   \\ 
 2454900.301029   &      24.72   &       4.72   &      -1.88   &       3.00   \\ 
 2454900.302532   &      25.29   &       5.12   &      -2.53   &       2.33   \\ 
 2454900.304032   &      28.38   &       4.61   &      -1.87   &       2.96   \\ 
 2454900.305535   &      28.30   &       4.98   &       1.33   &       6.12   \\ 
\hline
\end{tabular}

\end{table*}
 
\setcounter{table}{0}
\begin{table*}
\caption{Continued.}
\begin{tabular}{ccccc}
\hline
JD              &  RV	     & RV Error          &        RV     &      RV       \\
                 &[ms$^{-1}]$& [ms$^{-1}]$       & [ms$^{-1}]$   & [ms$^{-1}]$   \\
                 &  No corr  &                   & I corr        & A corr        \\
\hline
\multicolumn{5}{c}{Proxima Centauri ($\Gamma_{\rm ProxCen} = -22344.98 $ \ms) } \\
\hline
 2454900.307033   &      29.76   &       4.98   &       1.35   &       6.10   \\ 
 2454900.308533   &      25.95   &       4.68   &       2.90   &       7.61   \\ 
 2454900.310032   &      30.03   &       5.00   &      -0.82   &       3.85   \\ 
 2454900.311532   &      30.90   &       5.31   &       3.36   &       7.98   \\ 
 2454900.313032   &      26.97   &       4.13   &       4.33   &       8.90   \\ 
 2454900.314532   &      18.36   &       4.16   &       0.48   &       5.00   \\ 
 2454900.316038   &      24.23   &       4.48   &      -8.03   &      -3.57   \\ 
 2454900.317539   &      26.77   &       5.26   &      -2.05   &       2.35   \\ 
 2454900.319042   &      26.28   &       5.11   &       0.58   &       4.91   \\ 
 2454900.320542   &      27.62   &       5.11   &       0.19   &       4.45   \\ 
 2454900.322043   &      31.32   &       5.20   &       1.63   &       5.82   \\ 
 2454900.323542   &      29.51   &       5.67   &       5.44   &       9.56   \\ 
 2454900.325041   &      30.73   &       5.21   &       3.72   &       7.76   \\ 
 2454900.326542   &      27.28   &       4.87   &       5.05   &       9.00   \\ 
 2454900.328043   &      19.42   &       4.11   &       1.71   &       5.57   \\ 
 2454900.329542   &      17.95   &       3.98   &      -6.06   &      -2.28   \\ 
 2454900.331044   &      21.07   &       4.74   &      -7.44   &      -3.75   \\ 
 2454900.332541   &      23.46   &       4.80   &      -4.22   &      -0.62   \\ 
 2454900.334037   &      19.47   &       4.78   &      -1.72   &       1.77   \\ 
 2454900.335538   &      39.69   &       3.79   &      -5.61   &      -2.22   \\ 
 2454900.337035   &      26.45   &       5.07   &      14.71   &      18.00   \\ 
 2454900.338533   &      21.52   &       4.57   &       1.59   &       4.76   \\ 
 2454900.340030   &      26.91   &       4.64   &      -3.23   &      -0.18   \\ 
 2454900.341530   &      27.11   &       5.02   &       2.27   &       5.21   \\ 
 2454900.343027   &      22.89   &       5.45   &       2.55   &       5.37   \\ 
 2454900.344529   &      36.71   &       4.70   &      -1.55   &       1.14   \\ 
 2454900.346028   &      20.86   &       3.99   &      12.37   &      14.94   \\ 
 2454900.347525   &      20.47   &       4.58   &      -3.37   &      -0.93   \\ 
 2454900.349029   &      20.71   &       4.31   &      -3.66   &      -1.35   \\ 
 2454900.350530   &      24.89   &       4.57   &      -3.30   &      -1.14   \\ 
 2454900.352033   &      25.23   &       4.89   &       0.98   &       3.01   \\ 
 2454900.353534   &      27.66   &       5.01   &       1.43   &       3.31   \\ 
 2454900.355036   &      29.99   &       4.74   &       3.96   &       5.70   \\ 
 2454900.356534   &      24.30   &       4.51   &       6.40   &       7.99   \\ 
 2454900.358033   &      29.00   &       4.59   &       0.81   &       2.25   \\ 
 2454900.359529   &      23.27   &       4.66   &       5.62   &       6.90   \\ 
 2454900.361033   &      29.19   &       4.55   &       0.00   &       1.12   \\ 
 2454900.362531   &      17.42   &       4.00   &       6.03   &       6.98   \\ 
 2454900.364028   &      15.47   &       4.18   &      -5.63   &      -4.84   \\ 
 2454900.365530   &      14.90   &       4.12   &      -7.45   &      -6.84   \\ 
 2454900.367029   &      18.96   &       4.54   &      -7.91   &      -7.48   \\ 
 2454900.368529   &      23.75   &       4.41   &      -3.76   &      -3.50   \\ 
 2454900.370031   &      23.80   &       4.49   &       1.15   &       1.23   \\ 
 2454900.371533   &      16.53   &       4.52   &       1.32   &       1.21   \\ 
 2454900.373034   &      22.18   &       4.70   &      -5.84   &      -6.13   \\ 
 2454900.374529   &      23.47   &       5.21   &      -0.09   &      -0.57   \\ 
 2454900.376030   &      25.70   &       5.07   &       1.31   &       0.65   \\ 
 2454900.377529   &      16.35   &       4.57   &       3.66   &       2.79   \\ 
 2454900.379026   &      21.27   &       4.54   &      -5.60   &      -6.65   \\ 
 2454900.380525   &      17.30   &       4.59   &      -0.57   &      -1.83   \\ 
 2454900.382028   &      16.78   &       3.32   &      -4.41   &      -5.88   \\ 
 2454900.383531   &      21.45   &       4.03   &      -4.83   &      -6.50   \\ 
 2454900.385031   &      14.94   &       4.08   &      -0.05   &      -1.93   \\ 
 2454900.386529   &      22.67   &       4.11   &      -6.46   &      -8.55   \\ 
 2454900.388031   &      17.93   &       3.95   &       1.37   &      -0.93   \\ 
 2454900.389531   &      18.01   &       4.05   &      -3.25   &      -5.77   \\ 
 2454900.391030   &      24.50   &       4.60   &      -3.06   &      -5.80   \\ 
 2454900.392529   &      23.47   &       4.54   &       3.52   &       0.56   \\ 
 2454900.394030   &      24.73   &       4.05   &       2.61   &      -0.58   \\ 
\hline
\end{tabular}

\end{table*}
 
\setcounter{table}{0}
\begin{table*}
\caption{Continued.}
\begin{tabular}{ccccc}
\hline
JD              &  RV	     & RV Error          &        RV     &      RV       \\
                 &[ms$^{-1}]$& [ms$^{-1}]$       & [ms$^{-1}]$   & [ms$^{-1}]$   \\
                 &  No corr  &                   & I corr        & A corr        \\
\hline
\multicolumn{5}{c}{Proxima Centauri ($\Gamma_{\rm ProxCen} = -22344.98 $ \ms) } \\
\hline
 2454900.395533   &      18.80   &       4.07   &       3.96   &       0.55   \\ 
 2454900.397030   &      26.51   &       5.10   &      -1.83   &      -5.49   \\ 
 2454900.398529   &      19.49   &       4.48   &       5.97   &       2.08   \\ 
 2454900.400027   &      20.60   &       4.35   &      -0.94   &      -5.06   \\ 
 2454900.401529   &      14.70   &       4.09   &       0.27   &      -4.10   \\ 
 2454900.403029   &      20.03   &       3.18   &      -5.53   &     -10.13   \\ 
 2454900.404528   &      20.91   &       3.29   &      -0.09   &      -4.94   \\ 
 2454900.406026   &      23.33   &       3.55   &       0.88   &      -4.21   \\ 
 2454900.407524   &      18.27   &       3.20   &       3.41   &      -1.93   \\ 
 2454900.409026   &      13.12   &       3.45   &      -1.54   &      -7.14   \\ 
 2454900.410527   &      20.98   &       3.73   &      -6.58   &     -12.43   \\ 
 2454900.412027   &      16.86   &       2.61   &       1.36   &      -4.74   \\ 
 2454900.413525   &      17.44   &       4.32   &      -2.65   &      -9.01   \\ 
 2454900.415024   &      14.90   &       3.88   &      -1.98   &      -8.60   \\ 
 2454900.416524   &      16.05   &       3.92   &      -4.41   &     -11.30   \\ 
 2454900.418021   &      25.86   &       4.36   &      -3.18   &     -10.33   \\ 
 2454900.419523   &      21.91   &       4.16   &       6.73   &      -0.69   \\ 
 2454900.421022   &      17.45   &       3.73   &       2.89   &      -4.80   \\ 
 2454900.422519   &      22.23   &       3.99   &      -1.48   &      -9.44   \\ 
 2454900.424015   &      21.57   &       4.21   &       3.39   &      -4.84   \\ 
 2454900.425515   &      22.11   &       4.10   &       2.82   &      -5.68   \\ 
 2454900.426785   &      24.68   &       3.82   &       3.44   &      -5.34   \\ 
 2454902.126603   &      33.05   &       6.61   &       0.00   &       0.00   \\ 
 2454902.129258   &      31.04   &       6.27   &       2.86   &       5.31   \\ 
 2454902.131913   &      24.06   &       5.75   &       1.65   &       4.04   \\ 
 2454902.134571   &      16.01   &       6.22   &      -4.54   &      -2.21   \\ 
 2454902.137229   &      19.66   &       6.88   &     -11.79   &      -9.53   \\ 
 2454902.139885   &      28.96   &       5.74   &      -7.37   &      -5.17   \\ 
 2454902.142543   &      34.36   &       5.75   &       2.71   &       4.83   \\ 
 2454902.145201   &      26.18   &       4.87   &       8.87   &      10.93   \\ 
 2454902.147857   &      20.31   &       4.49   &       1.45   &       3.42   \\ 
 2454902.150522   &      26.96   &       4.75   &      -3.66   &      -1.76   \\ 
 2454902.153178   &      24.13   &       4.03   &       3.72   &       5.55   \\ 
 2454902.155834   &      16.90   &       3.65   &       1.63   &       3.38   \\ 
 2454902.158494   &      18.25   &       4.03   &      -4.86   &      -3.20   \\ 
 2454902.161152   &      20.63   &       4.39   &      -2.77   &      -1.19   \\ 
 2454902.163810   &      26.31   &       4.22   &       0.32   &       1.81   \\ 
 2454902.166468   &      27.95   &       3.98   &       6.71   &       8.12   \\ 
 2454902.169129   &      22.07   &       3.88   &       9.06   &      10.40   \\ 
 2454902.171790   &      13.96   &       3.18   &       3.88   &       5.13   \\ 
 2454902.174449   &      15.69   &       3.24   &      -3.53   &      -2.37   \\ 
 2454902.177107   &      11.78   &       3.39   &      -1.11   &      -0.04   \\ 
 2454902.179762   &      22.03   &       4.10   &      -4.34   &      -3.36   \\ 
 2454902.185073   &      22.81   &       3.52   &       6.60   &       7.50   \\ 
 2454902.187732   &      19.17   &       3.52   &       8.03   &       8.85   \\ 
 2454902.190384   &      17.70   &       3.53   &       5.06   &       5.79   \\ 
 2454902.193039   &      15.11   &       2.85   &       4.24   &       4.89   \\ 
 2454902.195693   &      13.83   &       2.28   &       2.31   &       2.88   \\ 
 2454902.198347   &      12.90   &       2.57   &       1.67   &       2.16   \\ 
 2454902.201003   &       5.11   &       2.48   &       1.38   &       1.78   \\ 
 2454902.203661   &      12.80   &       2.50   &      -5.78   &      -5.46   \\ 
 2454902.206322   &       7.87   &       2.61   &       2.54   &       2.78   \\ 
 2454902.208980   &       7.24   &       2.75   &      -1.77   &      -1.61   \\ 
 2454902.211640   &       0.08   &       2.50   &      -1.79   &      -1.70   \\ 
 2454902.214297   &       9.24   &       3.10   &      -8.35   &      -8.34   \\ 
 2454902.216955   &      14.56   &       3.61   &       1.40   &       1.34   \\ 
 2454902.219613   &       0.97   &       3.07   &       7.33   &       7.19   \\ 
 2454902.222269   &       4.40   &       2.90   &      -5.68   &      -5.89   \\ 
 2454902.224928   &       6.92   &       2.76   &      -1.67   &      -1.95   \\ 
\hline
\end{tabular}

\end{table*}
 
\setcounter{table}{0}
\begin{table*}
\caption{Continued.}
\begin{tabular}{ccccc}
\hline
JD              &  RV	     & RV Error          &        RV     &      RV       \\
                 &[ms$^{-1}]$& [ms$^{-1}]$       & [ms$^{-1}]$   & [ms$^{-1}]$   \\
                 &  No corr  &                   & I corr        & A corr        \\
\hline
\multicolumn{5}{c}{Proxima Centauri ($\Gamma_{\rm ProxCen} = -22344.98 $ \ms) } \\
\hline
 2454902.227584   &       8.47   &       2.35   &       1.42   &       1.07   \\ 
 2454902.230241   &       6.49   &       2.59   &       3.54   &       3.13   \\ 
 2454902.232605   &       3.72   &       2.67   &       2.11   &       1.64   \\ 
 2454902.234723   &       3.68   &       2.47   &      -0.12   &      -0.66   \\ 
 2454902.236801   &       1.24   &       2.31   &       0.28   &      -0.31   \\ 
 2454902.238878   &       1.89   &       2.56   &      -1.74   &      -2.37   \\ 
 2454902.240785   &       3.67   &       2.27   &      -0.68   &      -1.35   \\ 
 2454902.242534   &       2.60   &       2.57   &       1.52   &       0.80   \\ 
 2454902.244276   &       0.86   &       2.31   &       0.78   &       0.03   \\ 
 2454902.246016   &       7.56   &       4.58   &      -0.61   &      -1.39   \\ 
 2454902.247759   &      -0.36   &       2.55   &       6.42   &       5.60   \\ 
 2454902.249500   &      -1.60   &       2.18   &      -1.17   &      -2.02   \\ 
 2454902.251248   &      -4.03   &       2.24   &      -2.08   &      -2.96   \\ 
 2454902.252985   &      -7.44   &       2.33   &      -4.18   &      -5.09   \\ 
 2454902.254729   &      -3.64   &       2.59   &      -7.27   &      -8.20   \\ 
 2454902.256470   &      -2.85   &       2.55   &      -3.15   &      -4.10   \\ 
 2454902.258212   &      -1.44   &       2.58   &      -2.04   &      -3.03   \\ 
 2454902.270337   &       0.67   &       3.20   &      -0.32   &      -1.33   \\ 
 2454902.272073   &      -2.25   &       2.48   &       3.87   &       2.73   \\ 
 2454902.273808   &      -2.52   &       2.43   &       1.23   &       0.08   \\ 
 2454902.275548   &      -3.13   &       2.55   &       1.24   &       0.07   \\ 
 2454902.277298   &      -2.95   &       2.48   &       0.92   &      -0.26   \\ 
 2454902.279033   &      -1.37   &       2.73   &       1.37   &       0.17   \\ 
 2454902.280777   &      -2.80   &       2.44   &       3.22   &       2.02   \\ 
 2454902.282523   &      -0.22   &       2.55   &       2.06   &       0.85   \\ 
 2454902.284274   &      -7.32   &       2.59   &       4.90   &       3.69   \\ 
 2454902.286016   &      -3.69   &       2.74   &      -1.95   &      -3.17   \\ 
 2454902.287752   &     -16.50   &       2.72   &       1.94   &       0.71   \\ 
 2454902.289498   &     -13.52   &       2.70   &     -10.61   &     -11.84   \\ 
 2454902.291240   &      -4.51   &       2.82   &      -7.39   &      -8.62   \\ 
 2454902.292982   &     -15.89   &       3.28   &       1.87   &       0.64   \\ 
 2454902.294720   &      -8.83   &       2.98   &      -9.27   &     -10.50   \\ 
 2454902.296467   &      -6.91   &       3.81   &      -1.97   &      -3.20   \\ 
 2454902.298213   &     -12.70   &       4.44   &       0.19   &      -1.04   \\ 
 2454902.299962   &     -11.62   &       3.75   &      -5.39   &      -6.61   \\ 
 2454902.301708   &     -12.94   &       3.99   &      -4.07   &      -5.28   \\ 
 2454902.303454   &      -9.86   &       4.50   &      -5.17   &      -6.37   \\ 
 2454902.305194   &     -12.19   &       3.64   &      -1.88   &      -3.07   \\ 
 2454902.306931   &     -16.62   &       4.40   &      -3.99   &      -5.18   \\ 
 2454902.308668   &      -8.46   &       4.89   &      -8.21   &      -9.39   \\ 
 2454902.310403   &     -13.79   &       4.59   &       0.15   &      -1.01   \\ 
 2454902.312152   &     -10.64   &       4.90   &      -4.98   &      -6.12   \\ 
 2454902.313892   &      -9.62   &       4.73   &      -1.63   &      -2.77   \\ 
 2454902.315640   &      -4.56   &       5.32   &      -0.41   &      -1.53   \\ 
 2454902.317377   &     -15.27   &       5.27   &       4.84   &       3.74   \\ 
 2454902.319119   &     -13.13   &       3.95   &      -5.69   &      -6.77   \\ 
 2454902.320861   &     -12.23   &       4.95   &      -3.37   &      -4.42   \\ 
 2454902.322614   &     -17.73   &       5.08   &      -2.29   &      -3.32   \\ 
 2454902.324359   &     -13.19   &       3.80   &      -7.62   &      -8.63   \\ 
 2454902.326099   &      -8.31   &       4.09   &      -2.92   &      -3.90   \\ 
 2454902.327730   &     -10.80   &       3.98   &       2.13   &       1.17   \\ 
 2454902.329228   &     -13.21   &       4.34   &      -0.20   &      -1.13   \\ 
 2454902.330731   &     -19.33   &       4.00   &      -2.48   &      -3.38   \\ 
 2454902.332232   &      -8.97   &       3.91   &      -8.47   &      -9.35   \\ 
 2454902.333733   &      -6.23   &       3.86   &       2.02   &       1.17   \\ 
 2454902.335174   &     -13.85   &       3.13   &       4.89   &       4.07   \\ 
 2454902.336556   &     -14.38   &       3.77   &      -2.61   &      -3.40   \\ 
 2454902.337935   &     -12.89   &       3.38   &      -3.03   &      -3.80   \\ 
 2454902.339321   &      -4.54   &       4.17   &      -1.43   &      -2.17   \\ 
\hline
\end{tabular}

\end{table*}
 
\setcounter{table}{0}
\begin{table*}
\caption{Continued.}
\begin{tabular}{ccccc}
\hline
JD              &  RV	     & RV Error          &        RV     &      RV       \\
                 &[ms$^{-1}]$& [ms$^{-1}]$       & [ms$^{-1}]$   & [ms$^{-1}]$   \\
                 &  No corr  &                   & I corr        & A corr        \\
\hline
\multicolumn{5}{c}{Proxima Centauri ($\Gamma_{\rm ProxCen} = -22344.98 $ \ms) } \\
\hline
 2454902.340707   &     -15.83   &       4.26   &       7.02   &       6.31   \\ 
 2454902.342089   &     -18.69   &       3.55   &      -4.17   &      -4.85   \\ 
 2454902.343474   &     -23.75   &       3.38   &      -6.94   &      -7.58   \\ 
 2454902.344856   &      -9.56   &       3.25   &     -11.90   &     -12.51   \\ 
 2454902.346239   &      -5.76   &       3.11   &       2.39   &       1.81   \\ 
 2454902.347623   &     -11.68   &       3.16   &       6.28   &       5.73   \\ 
 2454902.349003   &     -17.19   &       3.24   &       0.44   &      -0.07   \\ 
 2454902.350390   &     -10.41   &       3.04   &      -4.98   &      -5.46   \\ 
 2454902.351776   &     -12.64   &       3.26   &       1.88   &       1.44   \\ 
 2454902.353159   &     -11.05   &       2.93   &      -0.27   &      -0.68   \\ 
 2454902.354547   &     -11.87   &       3.10   &       1.39   &       1.03   \\ 
 2454902.355931   &     -12.12   &       3.21   &       0.65   &       0.32   \\ 
 2454902.357315   &      -5.01   &       2.82   &       0.46   &       0.17   \\ 
 2454902.358700   &      -5.97   &       3.25   &       7.64   &       7.39   \\ 
 2454902.360086   &      -3.69   &       2.90   &       6.74   &       6.54   \\ 
 2454902.361473   &      -4.23   &       3.12   &       9.09   &       8.93   \\ 
 2454902.362860   &      -3.26   &       3.26   &       8.60   &       8.47   \\ 
 2454902.364245   &      -5.13   &       3.83   &       9.62   &       9.54   \\ 
 2454902.365630   &       0.26   &       3.19   &       7.81   &       7.78   \\ 
 2454902.367013   &      -6.34   &       3.15   &      13.25   &      13.26   \\ 
 2454902.368397   &      -5.76   &       3.37   &       6.70   &       6.76   \\ 
 2454902.369780   &       0.99   &       3.30   &       7.32   &       7.41   \\ 
 2454902.371165   &       0.79   &       3.26   &      14.12   &      14.26   \\ 
 2454902.372548   &       1.03   &       3.49   &      13.95   &      14.14   \\ 
 2454902.373934   &       0.54   &       3.67   &      14.22   &      14.46   \\ 
 2454902.375317   &      -3.24   &       3.62   &      13.77   &      14.05   \\ 
 2454902.376699   &       0.22   &       3.58   &      10.02   &      10.36   \\ 
 2454902.382023   &     -15.61   &       6.01   &      13.50   &      13.89   \\ 
 2454902.383406   &     -12.71   &       4.08   &      -2.26   &      -1.67   \\ 
 2454902.384788   &     -17.05   &       3.63   &       0.66   &       1.30   \\ 
 2454902.386175   &     -11.50   &       3.26   &      -3.67   &      -2.97   \\ 
 2454902.387557   &     -14.78   &       3.27   &       1.88   &       2.62   \\ 
 2454902.388938   &     -17.25   &       3.44   &      -1.40   &      -0.59   \\ 
 2454902.390325   &     -18.51   &       3.33   &      -3.86   &      -3.00   \\ 
 2454902.391707   &     -17.46   &       3.37   &      -5.13   &      -4.22   \\ 
 2454902.393089   &     -14.31   &       3.54   &      -4.09   &      -3.11   \\ 
 2454902.394476   &     -17.61   &       3.33   &      -0.95   &       0.08   \\ 
 2454902.395856   &     -20.69   &       3.67   &      -4.26   &      -3.16   \\ 
 2454902.397237   &     -17.11   &       3.17   &      -7.36   &      -6.21   \\ 
 2454902.398621   &     -13.22   &       3.50   &      -3.79   &      -2.58   \\ 
 2454902.400008   &     -11.37   &       3.18   &       0.07   &       1.34   \\ 
 2454902.401390   &     -16.99   &       3.38   &       1.90   &       3.23   \\ 
 2454902.402773   &      -9.81   &       3.72   &      -3.75   &      -2.36   \\ 
 2454902.404156   &     -19.08   &       3.39   &       3.40   &       4.86   \\ 
 2454902.405540   &     -22.88   &       3.64   &      -5.90   &      -4.38   \\ 
 2454902.406924   &     -19.98   &       3.37   &      -9.74   &      -8.17   \\ 
 2454902.408309   &     -14.75   &       3.38   &      -6.88   &      -5.24   \\ 
 2454902.409690   &     -19.41   &       3.39   &      -1.69   &       0.01   \\ 
 2454902.411071   &     -21.46   &       3.59   &      -6.40   &      -4.63   \\ 
 2454902.412455   &     -16.93   &       3.82   &      -8.50   &      -6.67   \\ 
 2454902.413836   &     -18.46   &       3.27   &      -4.02   &      -2.12   \\ 
 2454902.415221   &     -13.46   &       3.53   &      -5.61   &      -3.65   \\ 
 2454902.416149   &     -18.24   &       4.64   &      -0.67   &       1.37   \\ 
 2454904.125571   &       2.65   &       8.29   &       0.00   &       0.00   \\ 
 2454904.129383   &      -0.44   &       8.47   &      -9.04   &       4.64   \\ 
 2454904.133195   &       3.92   &       8.24   &     -10.37   &       2.44   \\ 
 2454904.137009   &      10.42   &       8.16   &      -4.27   &       7.69   \\ 
 2454904.140480   &       1.46   &       8.26   &       3.93   &      15.05   \\ 
\hline
\end{tabular}

\end{table*}
 
\setcounter{table}{0}
\begin{table*}
\caption{Continued.}
\begin{tabular}{ccccc}
\hline
JD              &  RV	     & RV Error          &        RV     &      RV       \\
                 &[ms$^{-1}]$& [ms$^{-1}]$       & [ms$^{-1}]$   & [ms$^{-1}]$   \\
                 &  No corr  &                   & I corr        & A corr        \\
\hline
\multicolumn{5}{c}{Proxima Centauri ($\Gamma_{\rm ProxCen} = -22344.98 $ \ms) } \\
\hline
 2454904.143603   &       5.21   &       8.31   &      -3.44   &       6.89   \\ 
 2454904.146742   &       2.80   &       8.53   &       1.66   &      11.34   \\ 
 2454904.149860   &       3.40   &       8.37   &       0.57   &       9.61   \\ 
 2454904.152999   &       3.53   &       8.27   &       2.46   &      10.88   \\ 
 2454904.156117   &      11.14   &       8.40   &       3.87   &      11.68   \\ 
 2454904.159235   &      -8.33   &       9.65   &      12.74   &      19.95   \\ 
 2454904.162353   &      -6.52   &       8.82   &      -5.49   &       1.13   \\ 
 2454904.168362   &     -11.38   &       5.21   &      -2.47   &       3.59   \\ 
 2454904.171021   &     -16.80   &       6.17   &      -4.99   &      -0.02   \\ 
 2454904.173680   &     -12.38   &       5.28   &      -9.42   &      -4.90   \\ 
 2454904.176046   &      -6.71   &       5.05   &      -4.03   &       0.06   \\ 
 2454904.178121   &      -0.86   &       6.57   &       2.56   &       6.23   \\ 
 2454904.180196   &      -6.80   &       5.50   &       9.16   &      12.49   \\ 
 2454904.182099   &      -5.26   &       5.29   &       3.94   &       6.96   \\ 
 2454904.183837   &      -8.69   &       6.11   &       6.20   &       8.90   \\ 
 2454904.185583   &      -2.41   &       6.21   &       3.37   &       5.82   \\ 
 2454904.187322   &      -9.86   &       6.76   &      10.24   &      12.43   \\ 
 2454904.189064   &     -13.40   &       7.07   &       3.38   &       5.32   \\ 
 2454904.190803   &      -3.62   &       6.95   &       0.43   &       2.12   \\ 
 2454904.192539   &      -3.40   &       6.70   &      10.79   &      12.24   \\ 
 2454904.194277   &      -7.03   &       6.91   &      11.58   &      12.80   \\ 
 2454904.196016   &     -10.79   &       7.10   &       8.52   &       9.50   \\ 
 2454904.197752   &     -12.24   &       5.89   &       5.33   &       6.09   \\ 
 2454904.199500   &     -12.09   &       7.43   &       4.44   &       4.96   \\ 
 2454904.201244   &      -9.33   &       6.80   &       5.15   &       5.45   \\ 
 2454904.202987   &     -10.78   &       7.61   &       8.45   &       8.54   \\ 
 2454904.204614   &     -16.28   &       6.87   &       7.55   &       7.43   \\ 
 2454904.206113   &     -19.01   &       6.57   &       2.57   &       2.25   \\ 
 2454904.207615   &     -11.09   &       6.73   &       0.28   &      -0.21   \\ 
 2454904.212117   &     -32.75   &       8.32   &       8.67   &       8.00   \\ 
 2454904.213617   &     -33.07   &       7.60   &     -11.66   &     -12.82   \\ 
 2454904.215119   &     -32.89   &       6.72   &     -11.53   &     -12.85   \\ 
 2454904.216616   &     -28.62   &       7.64   &     -10.92   &     -12.39   \\ 
 2454904.218113   &     -12.62   &       6.15   &      -6.23   &      -7.84   \\ 
 2454904.221645   &     -21.89   &       6.74   &      10.27   &       8.49   \\ 
 2454904.223146   &     -23.84   &       6.28   &       1.91   &      -0.19   \\ 
 2454904.224645   &     -27.45   &       8.07   &       0.39   &      -1.84   \\ 
 2454904.226144   &     -29.45   &       6.49   &      -2.83   &      -5.19   \\ 
 2454904.227641   &     -28.60   &       6.01   &      -4.43   &      -6.92   \\ 
 2454904.229145   &     -33.08   &       6.23   &      -3.18   &      -5.80   \\ 
 2454904.230646   &     -31.46   &       6.64   &      -7.27   &     -10.00   \\ 
 2454904.232148   &     -26.63   &       6.61   &      -5.26   &      -8.11   \\ 
 2454904.233646   &     -24.24   &       6.99   &      -0.04   &      -3.00   \\ 
 2454904.235148   &     -31.91   &       6.11   &       2.74   &      -0.34   \\ 
 2454904.236650   &     -25.18   &       4.53   &      -4.55   &      -7.74   \\ 
 2454904.238151   &     -36.69   &       4.83   &       2.55   &      -0.74   \\ 
 2454904.239654   &     -36.28   &       5.60   &      -8.60   &     -11.99   \\ 
 2454904.241155   &     -33.87   &       5.72   &      -7.82   &     -11.30   \\ 
 2454904.242658   &     -35.41   &       4.91   &      -5.04   &      -8.62   \\ 
 2454904.244161   &     -35.85   &       5.20   &      -6.22   &      -9.89   \\ 
 2454904.245663   &     -33.20   &       4.42   &      -6.32   &     -10.07   \\ 
 2454904.247164   &     -38.42   &       4.77   &      -3.30   &      -7.14   \\ 
 2454904.248663   &     -27.30   &       4.15   &      -8.19   &     -12.10   \\ 
 2454904.250166   &     -30.55   &       4.24   &       3.27   &      -0.72   \\ 
 2454904.251664   &     -34.97   &       4.34   &       0.37   &      -3.70   \\ 
 2454904.253163   &     -38.68   &       4.30   &      -3.73   &      -7.86   \\ 
 2454904.254664   &     -28.03   &       5.09   &      -7.11   &     -11.31   \\ 
 2454904.256162   &     -18.97   &       4.07   &       3.86   &      -0.40   \\ 
 2454904.257662   &     -21.51   &       4.37   &      13.25   &       8.93   \\ 
\hline
\end{tabular}

\end{table*}
 
\setcounter{table}{0}
\begin{table*}
\caption{Continued.}
\begin{tabular}{ccccc}
\hline
JD              &  RV	     & RV Error          &        RV     &      RV       \\
                 &[ms$^{-1}]$& [ms$^{-1}]$       & [ms$^{-1}]$   & [ms$^{-1}]$   \\
                 &  No corr  &                   & I corr        & A corr        \\
\hline
\multicolumn{5}{c}{Proxima Centauri ($\Gamma_{\rm ProxCen} = -22344.98 $ \ms) } \\
\hline
 2454904.259165   &     -23.78   &       4.37   &      11.01   &       6.64   \\ 
 2454904.260693   &     -34.13   &       5.76   &       9.06   &       4.64   \\ 
 2454904.262196   &     -27.92   &       5.11   &      -0.98   &      -5.45   \\ 
 2454904.263697   &     -29.30   &       4.78   &       5.53   &       1.02   \\ 
 2454904.265195   &     -18.86   &       4.70   &       4.45   &      -0.10   \\ 
 2454904.266693   &     -30.26   &       4.53   &      15.18   &      10.59   \\ 
 2454904.268190   &     -34.08   &       5.68   &       4.09   &      -0.53   \\ 
 2454904.269690   &     -38.42   &       5.37   &       0.54   &      -4.11   \\ 
 2454904.271188   &     -35.96   &       5.15   &      -3.52   &      -8.20   \\ 
 2454904.272686   &     -42.16   &       5.30   &      -0.79   &      -5.49   \\ 
 2454904.274185   &     -41.16   &       7.69   &      -6.72   &     -11.44   \\ 
 2454904.275685   &     -47.10   &       7.89   &      -5.45   &     -10.18   \\ 
 2454904.277185   &     -42.60   &      12.87   &     -11.12   &     -15.87   \\ 
 2454904.278687   &     -40.96   &       9.02   &      -6.37   &     -11.12   \\ 
 2454904.280186   &     -36.82   &       8.93   &      -4.46   &      -9.22   \\ 
 2454904.281690   &     -38.16   &      13.89   &      -0.08   &      -4.84   \\ 
 2454904.283190   &     -36.04   &       8.51   &      -1.16   &      -5.92   \\ 
 2454904.284686   &     -40.72   &       8.61   &       1.20   &      -3.56   \\ 
 2454904.286188   &     -37.83   &       8.29   &      -3.25   &      -8.00   \\ 
 2454904.287690   &     -36.67   &       7.86   &      -0.13   &      -4.86   \\ 
 2454904.289194   &     -40.66   &       8.40   &       1.26   &      -3.45   \\ 
 2454904.290697   &     -47.47   &       6.54   &      -2.52   &      -7.21   \\ 
 2454904.292198   &     -45.22   &       6.22   &      -9.09   &     -13.77   \\ 
 2454904.293696   &     -40.58   &       5.95   &      -6.63   &     -11.28   \\ 
 2454904.295194   &     -47.68   &       6.17   &      -1.79   &      -6.40   \\ 
 2454904.298196   &     -34.35   &       4.93   &      -8.67   &     -13.26   \\ 
 2454904.299695   &     -37.24   &       4.80   &       5.04   &       0.54   \\ 
 2454904.301196   &     -41.92   &       5.57   &       2.35   &      -2.11   \\ 
 2454904.302695   &     -47.51   &       8.46   &      -2.14   &      -6.55   \\ 
 2454904.304194   &     -47.88   &       7.77   &      -7.54   &     -11.90   \\ 
 2454904.305694   &     -44.81   &       7.62   &      -7.74   &     -12.05   \\ 
 2454904.307195   &     -35.30   &       7.33   &      -4.49   &      -8.75   \\ 
 2454904.308694   &     -48.91   &       6.74   &       5.19   &       1.00   \\ 
 2454904.310196   &     -44.49   &       7.64   &      -8.26   &     -12.39   \\ 
 2454904.311695   &     -47.00   &       6.00   &      -3.67   &      -7.74   \\ 
 2454904.313195   &     -42.54   &       5.35   &      -6.03   &     -10.02   \\ 
 2454904.314697   &     -44.21   &       5.79   &      -1.42   &      -5.34   \\ 
 2454904.316198   &     -23.45   &       6.41   &      -2.95   &      -6.78   \\ 
 2454904.317701   &     -38.44   &       6.34   &      17.96   &      14.21   \\ 
 2454904.319204   &     -40.51   &       5.35   &       3.10   &      -0.57   \\ 
 2454904.320705   &     -32.06   &       4.99   &       1.17   &      -2.41   \\ 
 2454904.322203   &     -39.31   &       5.11   &       9.73   &       6.25   \\ 
 2454904.323705   &     -41.10   &       7.64   &       2.60   &      -0.79   \\ 
 2454904.325202   &     -39.15   &       7.90   &       0.94   &      -2.36   \\ 
 2454904.326699   &     -41.71   &       6.57   &       3.00   &      -0.19   \\ 
 2454904.328198   &     -46.54   &       7.73   &       0.55   &      -2.54   \\ 
 2454904.329700   &     -35.74   &       6.12   &      -4.17   &      -7.15   \\ 
 2454904.331201   &     -43.24   &       6.07   &       6.73   &       3.86   \\ 
 2454904.332698   &     -40.61   &       5.27   &      -0.69   &      -3.44   \\ 
 2454904.334200   &     -35.91   &       5.66   &       2.05   &      -0.59   \\ 
 2454904.335698   &     -39.84   &       5.87   &       6.83   &       4.31   \\ 
 2454904.337198   &     -44.96   &       5.52   &       2.98   &       0.59   \\ 
 2454904.338696   &     -35.15   &       5.23   &      -2.06   &      -4.33   \\ 
 2454904.340199   &     -42.19   &       6.26   &       7.81   &       5.68   \\ 
 2454904.341698   &     -45.59   &       8.56   &       0.83   &      -1.16   \\ 
 2454904.343196   &     -40.51   &       6.83   &      -2.51   &      -4.36   \\ 
 2454904.344695   &     -35.16   &       8.27   &       2.63   &       0.92   \\ 
\hline
\end{tabular}

\end{table*}
 
\setcounter{table}{0}
\begin{table*}
\caption{Continued.}
\begin{tabular}{ccccc}
\hline
JD              &  RV	     & RV Error          &        RV     &      RV       \\
                 &[ms$^{-1}]$& [ms$^{-1}]$       & [ms$^{-1}]$   & [ms$^{-1}]$   \\
                 &  No corr  &                   & I corr        & A corr        \\
\hline
\multicolumn{5}{c}{Proxima Centauri ($\Gamma_{\rm ProxCen} = -22344.98 $ \ms) } \\
\hline
 2454904.346195   &     -47.53   &       7.27   &       8.03   &       6.46   \\ 
 2454904.347695   &     -46.04   &       7.92   &      -4.29   &      -5.71   \\ 
 2454904.349197   &     -47.62   &       6.46   &      -2.76   &      -4.03   \\ 
 2454904.350700   &     -44.62   &       6.25   &      -4.30   &      -5.42   \\ 
 2454904.352198   &     -43.40   &       6.39   &      -1.27   &      -2.23   \\ 
 2454904.353698   &     -44.25   &       5.92   &      -0.01   &      -0.82   \\ 
 2454904.355195   &     -42.79   &       6.10   &      -0.87   &      -1.50   \\ 
 2454904.356699   &     -46.19   &       5.82   &       0.61   &       0.14   \\ 
 2454904.358198   &     -35.71   &       5.22   &      -2.76   &      -3.06   \\ 
 2454904.359699   &     -48.57   &       6.21   &       7.72   &       7.59   \\ 
 2454904.361201   &     -44.08   &      10.51   &      -5.13   &      -5.09   \\ 
 2454904.362702   &     -42.37   &       7.63   &      -0.66   &      -0.43   \\ 
 2454904.364204   &     -43.55   &       8.88   &       1.04   &       1.45   \\ 
 2454904.365702   &     -46.64   &       7.47   &      -0.14   &       0.45   \\ 
 2454904.367200   &     -42.91   &       7.46   &      -3.26   &      -2.48   \\ 
 2454904.368703   &     -50.02   &       5.99   &       0.46   &       1.43   \\ 
 2454904.370207   &     -39.24   &       5.55   &      -6.68   &      -5.52   \\ 
 2454904.371704   &     -41.52   &       5.47   &       4.07   &       5.43   \\ 
 2454904.373200   &     -38.82   &       5.95   &       1.74   &       3.30   \\ 
 2454904.374701   &     -42.23   &       5.77   &       4.40   &       6.17   \\ 
 2454904.376200   &     -38.00   &       5.31   &       0.95   &       2.92   \\ 
 2454904.377701   &     -34.46   &       5.99   &       5.12   &       7.30   \\ 
 2454904.379200   &     -41.17   &       5.17   &       8.59   &      10.98   \\ 
 2454904.380703   &     -34.71   &       9.18   &       1.83   &       4.43   \\ 
 2454904.382198   &     -40.94   &       8.52   &       8.21   &      11.03   \\ 
 2454904.383698   &     -45.30   &       7.23   &       1.91   &       4.95   \\ 
 2454904.385200   &     -46.01   &       7.80   &      -2.53   &       0.73   \\ 
 2454904.386696   &     -45.85   &       8.57   &      -3.31   &       0.18   \\ 
 2454904.388221   &     -41.80   &       6.14   &      -3.26   &       0.46   \\ 
 2454904.389721   &     -46.79   &       6.55   &       0.70   &       4.66   \\ 
 2454904.391222   &     -38.18   &       6.36   &      -4.38   &      -0.19   \\ 
 2454904.392720   &     -46.95   &       7.29   &       4.12   &       8.55   \\ 
 2454904.394218   &     -49.88   &       6.61   &      -4.76   &      -0.10   \\ 
 2454904.395717   &     -42.59   &       7.42   &      -7.80   &      -2.89   \\ 
 2454904.397217   &     -40.04   &       5.74   &      -0.63   &       4.52   \\ 
 2454904.398718   &     -40.72   &       6.05   &       1.79   &       7.19   \\ 
 2454904.400220   &     -31.25   &       5.46   &       0.96   &       6.62   \\ 
 2454904.401720   &     -36.27   &       5.66   &      10.31   &      16.22   \\ 
 2454904.403218   &     -42.01   &       7.45   &       5.15   &      11.31   \\ 
 2454904.404719   &     -38.31   &       8.50   &      -0.74   &       5.68   \\ 
 2454904.406218   &     -36.81   &       7.99   &       2.80   &       9.48   \\ 
 2454904.407718   &     -50.61   &       7.69   &       4.16   &      11.10   \\ 
 2454904.409249   &     -48.71   &       7.32   &      -9.81   &      -2.59   \\ 
 2454904.410747   &     -40.68   &       8.22   &      -8.08   &      -0.60   \\ 
 2454904.412244   &     -39.73   &       8.12   &      -0.23   &       7.54   \\ 
\hline
\end{tabular}

\end{table*}

\setcounter{table}{1}
\begin{table*}
\caption{Observation times and radial velocities for all M5V\,-\,M7V ROPS targets deconvolved with the GJ 1061 line list. The $\Gamma_*$ velocity indicated in each case must be added to the velocities to transform them into the reference frame of GJ 1061 (from which the deconvolution template was derived). Columns 1\,-\,6 are Julian date, raw radial velocity with $\Gamma_*$ subtracted (No corr), propagated error, stellar line bisector corrected velocity (L corr), telluric bisector corrected velocity (T corr) and stellar line minus telluric line bisector corrected velocity (L-T corr).}
\begin{tabular}{cccccc}
\hline
 JD              &  RV	     & RV Error          &        RV     &      RV       &       RV    \\
                 &[ms$^{-1}]$& [ms$^{-1}]$       & [ms$^{-1}]$   & [ms$^{-1}]$   & [ms$^{-1}]$ \\
                 &  No corr  &                   & L corr        & T corr        & L-T corr    \\
\hline
\multicolumn{6}{c}{GJ 3076 ( $\Gamma_{\rm GJ3076} = 2449.11 $ \ms) } \\
\hline

 2456131.834179   &   -9.18   &           11.55   &      -49.64   &      -41.32   &       39.43   \\ 
 2456132.832890   &  137.39   &           23.09   &      116.00   &       52.32   &        1.98   \\ 
 2456134.859510   &  -23.93   &           18.38   &       27.71   &       62.41   &       21.24   \\ 
 2456137.817710   & -104.28   &           35.85   &      -94.08   &      -73.41   &      -62.65   \\ 

\hline
\multicolumn{6}{c}{GJ 1002 ( $\Gamma_{\rm GJ1002} = -40733.57 $ \ms) } \\
\hline

 2456131.766112   &   11.95   &            6.32   &       -1.05   &      -12.68   &       20.81   \\ 
 2456132.763310   &  -38.07   &            4.04   &       -4.50   &       -2.18   &      -33.53   \\ 
 2456134.782990   &   -4.52   &            5.44   &        7.19   &       -2.81   &        7.76   \\ 
 2456137.825820   &   30.64   &            9.62   &       -1.65   &       17.67   &        4.96   \\ 

\hline
\multicolumn{6}{c}{GJ 1061 ( $\Gamma_{\rm GJ1061} = 14499.07 $ \ms) } \\
\hline

 2456131.905029   &    5.54   &           11.48   &        2.98   &        3.61   &        2.39   \\ 
 2456132.912720   &    0.58   &            5.16   &        2.08   &        1.01   &        3.62   \\ 
 2456132.917280   &   -1.58   &            4.87   &       -1.92   &       -1.98   &       -1.74   \\ 
 2456134.920640   &    1.47   &            7.44   &       -2.25   &       -2.21   &       -1.64   \\ 
 2456137.897020   &   -6.01   &            6.80   &       -0.89   &       -0.43   &       -2.63   \\ 

\hline
\multicolumn{6}{c}{LP 759-25 ( $\Gamma_{\rm LP759-25} = 15661.08 $ \ms) } \\
\hline

 2456131.751882   &  -81.56   &           18.58   &       18.70   &      -72.16   &       21.35   \\ 
 2456132.750070   &  -75.07   &            9.53   &      -76.08   &       34.48   &      -40.98   \\ 
 2456134.768600   &  147.79   &           14.79   &      104.11   &       81.46   &       83.02   \\ 
 2456137.709730   &    8.83   &           14.24   &      -46.73   &      -43.78   &      -63.39   \\ 

\hline
\multicolumn{6}{c}{GJ 3146 ( $\Gamma_{\rm GJ3146} = 14908.17 $ \ms) } \\
\hline

 2456131.894923   &  -64.05   &           14.20   &      -44.64   &      -40.08   &       -9.85   \\ 
 2456132.904030   &   59.38   &            8.88   &       66.48   &        8.83   &       -2.06   \\ 
 2456134.937430   &  -84.83   &           20.23   &      -10.32   &      -76.91   &        7.99   \\ 
 2456137.887810   &   89.51   &           26.83   &      -11.52   &      108.15   &        3.91   \\ 

\hline
\multicolumn{6}{c}{GJ 3128 ( $\Gamma_{\rm GJ3128} = 20020.94 $ \ms) } \\
\hline

 2456131.842671   &   33.86   &           13.54   &        2.29   &       35.22   &        7.38   \\ 
 2456132.840830   &  -15.79   &            6.95   &        0.29   &      -15.09   &       -3.64   \\ 
 2456134.869770   &    1.77   &            7.54   &       12.66   &       -4.27   &       16.28   \\ 
 2456137.835510   &  -19.85   &            7.97   &      -15.23   &      -15.86   &      -20.02   \\ 

\hline
\multicolumn{6}{c}{GJ 4281 ( $\Gamma_{\rm GJ4281} = -7408.90 $ \ms) } \\
\hline

 2456131.735050   &   40.61   &           10.94   &       14.92   &        0.24   &       -4.93   \\ 
 2456132.734400   &    9.37   &            4.86   &      -13.31   &        2.08   &       21.38   \\ 
 2456134.741200   &   -2.04   &            5.68   &       -3.18   &       -2.78   &       -2.13   \\ 
 2456137.694200   &  -47.94   &           12.84   &        1.57   &        0.47   &      -14.32   \\ 

\hline
\multicolumn{6}{c}{SO J025300.5+165258 ( $\Gamma_{\rm SOJ0253+1652} = 64010.47 $ \ms) } \\
\hline

 2456131.912948   &    2.39   &           13.48   &       -6.51   &        3.37   &       -3.83   \\ 
 2456132.925100   &  -20.07   &            4.53   &      -14.29   &      -18.20   &      -17.08   \\ 
 2456134.929120   &   16.80   &            9.96   &       10.94   &        4.94   &       17.97   \\ 
 2456137.905420   &    0.87   &           10.56   &        9.86   &        9.89   &        2.94   \\ 

\hline
 \end{tabular}

\protect\label{tab:appendix_table_gj1061}

\end{table*}

\setcounter{table}{2}
\begin{table*}
\caption{Observation times and velocities for the M7.5V\,-\,M9V targets deconvolved with the LHS 132 line list.}
\begin{tabular}{cccccc}
\hline
JD              &  RV	     & RV Error          &        RV     &      RV       &       RV    \\
                 &[ms$^{-1}]$& [ms$^{-1}]$       & [ms$^{-1}]$   & [ms$^{-1}]$   & [ms$^{-1}]$ \\
                 &  No corr  &                   & L corr        & T corr        & L-T corr    \\
\hline
\multicolumn{6}{c}{LP 888-18 ( $\Gamma_{\rm LP888-18} = 25171.22 $ \ms) } \\
\hline

 2456131.858119   &  -61.82   &           26.26   &      -51.26   &      -30.83   &      -38.66   \\ 
 2456132.864200   &   21.77   &            6.11   &       13.88   &      -21.16   &      -13.01   \\ 
 2456134.885720   &   43.01   &            9.99   &        8.22   &       37.02   &       51.60   \\ 
 2456137.851140   &   -2.97   &           11.35   &       29.16   &       14.97   &        0.07   \\ 

\hline
\multicolumn{6}{c}{LHS 132 ( $\Gamma_{\rm LHS132} = 18661.17 $ \ms) } \\
\hline

 2456131.803782   &    0.03   &           15.84   &        1.97   &       -3.04   &       -5.82   \\ 
 2456132.799410   &   13.05   &            6.64   &       15.07   &       10.81   &       13.53   \\ 
 2456134.824210   &   -4.33   &            8.73   &       -2.32   &       -7.33   &       -4.63   \\ 
 2456137.784340   &  -16.72   &           13.51   &      -14.72   &       -0.44   &       -3.08   \\ 

\hline
\multicolumn{6}{c}{2MASS J23062928-0502285 ( $\Gamma_{\rm 2MJ23-05} = -51688.03 $ \ms) } \\
\hline

 2456131.716866   &  -40.55   &           15.19   &      -14.89   &       -6.83   &        0.17   \\ 
 2456132.714250   &   -1.12   &            6.58   &       12.99   &      -19.35   &      -13.81   \\ 
 2456134.721330   &   19.56   &            9.64   &       11.21   &        6.09   &        3.93   \\ 
 2456137.728420   &   22.11   &           19.30   &       -9.31   &       20.09   &        9.71   \\ 

\hline
\multicolumn{6}{c}{LHS 1367 ( $\Gamma_{\rm LHS1367} = 657.43 $ \ms) } \\
\hline

 2456131.821169   &   -5.43   &            8.95   &      -17.23   &       -7.49   &       -2.41   \\ 
 2456132.818410   &   22.57   &            4.65   &        7.97   &        0.87   &       18.50   \\ 
 2456134.844390   &   12.19   &            6.02   &       16.77   &       22.06   &       14.71   \\ 
 2456137.803370   &  -29.32   &            7.56   &       -7.51   &      -15.43   &      -30.80   \\ 

\hline
\multicolumn{6}{c}{LP412-31 ( $\Gamma_{\rm LP412-31} = 46612.38 $ \ms) } \\
\hline

 2456131.925114   &  152.74   &           12.83   &       13.71   &      192.41   &      -69.72   \\ 
 2456132.934000   &  139.53   &           10.68   &      215.47   &       88.73   &      138.13   \\ 
 2456137.917920   & -292.27   &           17.15   &     -229.18   &     -281.14   &      -68.42   \\ 

\hline
\multicolumn{6}{c}{2MASS J23312174-2749500 ( $\Gamma_{\rm 2MJ23-27} = 261.26 $ \ms) } \\
\hline

 2456131.782437   &  -53.61   &           11.25   &      -54.65   &      -27.25   &      -20.69   \\ 
 2456132.779070   &   21.80   &            6.06   &       21.11   &       30.22   &       30.79   \\ 
 2456134.801980   &    3.71   &            7.50   &       11.93   &      -23.18   &        0.40   \\ 
 2456137.761310   &   28.10   &           11.44   &       21.62   &       20.20   &      -10.50   \\ 

\hline
\multicolumn{6}{c}{2MASS J03341218-4953322 ( $\Gamma_{\rm 2MJ03-49} = 73732.21 $ \ms) } \\
\hline

 2456131.879267   &   -2.05   &            7.31   &       -1.64   &       -8.18   &       -9.37   \\ 
 2456132.886170   &  -14.65   &            5.18   &       -5.81   &       -2.95   &       -4.68   \\ 
 2456134.906790   &    5.63   &            7.63   &        9.08   &        7.29   &        6.13   \\ 
 2456137.872580   &   11.07   &            8.16   &       -1.63   &        3.84   &        7.92   \\ 

\hline
                 
\end{tabular}
\protect\label{tab:appendix_table_lhs132}
\end{table*}

\clearpage
\setcounter{table}{3}
\begin{table*}
\begin{center}
\caption{Radial velocities for GJ 1061 and GJ 1002 derived using {\sc terra} (see \S \ref{section:discussion}). The spectra were taken from the {\sc eso} {\sc harps} archive.}
\begin{tabular}{ccc}
\hline
JD              &  RV	      & RV Error           \\
                & [ms$^{-1}]$ & [ms$^{-1}]$        \\
                &             &                    \\
\hline
\multicolumn{3}{c}{GJ 1061} \\
\hline
2452985.713012   &    0.35    &      1.09 \\      
2452996.737269   &    0.00    &      1.23 \\      
2453337.748816   &   -3.51    &      0.85 \\      
2454341.868575   &   -3.18    &      0.93 \\      
\hline
\multicolumn{3}{c}{GJ 1002} \\
\hline
2453336.603252    &    3.01    &     1.66 \\       
2453918.940758    &    0.00    &     1.69 \\     
2454048.614913    &   -0.61    &     1.64 \\      
2454800.563001    &   -2.58    &     1.42 \\      
\hline
\end{tabular}
\end{center}
\protect\label{tab:appendix_table_terra}
\end{table*}

\protect\label{lastpage}
\end{document}